\documentclass[sigconf, screen]{acmart}

\usepackage{multirow}
\usepackage{algorithm, algorithmic}
\usepackage{makecell}
\usepackage{enumitem}
\usepackage{listings}

\renewcommand\footnotetextcopyrightpermission[1]{} 
\setcopyright{none}

\acmDOI{}

\acmISBN{}

\begin{document}
\title{RailX: A Flexible, Scalable, and Low-Cost Network Architecture for Hyper-Scale LLM Training Systems}

\author{Yinxiao Feng}
\affiliation{%
  \institution{Tsinghua University}
}
\email{fyx20@mails.tsinghua.edu.cn}

\author{Tiancheng Chen}
\affiliation{%
  \institution{ETH Zurich}
}
\email{tiancheng.chen@inf.ethz.ch}

\author{Yuchen Wei}
\affiliation{%
  \institution{Tsinghua University}
}
\email{weiyc22@mails.tsinghua.edu.cn}

\author{Siyuan Shen}
\affiliation{%
  \institution{ETH Zurich}
}
\email{siyuan.shen@inf.ethz.ch}

\author{Shiju Wang}
\affiliation{%
  \institution{Beihang University}
}
\email{21373455@buaa.edu.cn}

\author{Wei Li}
\affiliation{%
  \institution{Tsinghua University}
}
\email{liw21@mails.tsinghua.edu.cn}

\author{Kaisheng Ma}
\authornote{Corresponding Author}
\affiliation{%
  \institution{Tsinghua University}
}
\email{kaisheng@mail.tsinghua.edu.cn}

\author{Torsten Hoefler}
\affiliation{%
  \institution{ETH Zurich}
}
\email{torsten.hoefler@inf.ethz.ch}

\begin{abstract}
    Increasingly large AI workloads are calling for hyper-scale infrastructure; however, traditional interconnection network architecture is neither scalable nor cost-effective enough. Tree-based topologies such as the \textit{Rail-optimized} network are extremely expensive, while direct topologies such as \textit{Torus} have insufficient bisection bandwidth and flexibility. In this paper, we propose \textit{RailX}, a reconfigurable network architecture based on intra-node direct connectivity and inter-node circuit switching. Nodes and optical switches are physically 2D-organized, achieving better scalability than existing centralized circuit switching networks. We propose a novel interconnection method based on \textit{Hamiltonian Decomposition} theory to organize separate rail-based rings into \textit{all-to-all} topology, simultaneously optimizing ring-collective and all-to-all communication. More than $100$K chips with hyper bandwidth can be interconnected with a flat switching layer, and the diameter is only $2\sim4$ inter-node hops. The network cost per injection/All-Reduce bandwidth of \textit{RailX} is less than $10\%$ of the Fat-Tree, and the cost per bisection/All-to-All bandwidth is less than $50\%$ of the Fat-Tree. Specifically, only $\sim$\$$1.3$B is required to interconnect 200K chips with 1.8TB bandwidth. \textit{RailX} can also be used in the ML-as-a-service (MLaaS) scenario, where single or multiple training workloads with various shapes, scales, and parallelism strategies can be flexibly mapped, and failures can be worked around.
\end{abstract}

\maketitle

\pagestyle{plain}

\section{Introduction}
In recent years, AI workloads have become increasingly large, growing much faster than the hardware~\cite{Kaplan_ScalingLawsNeural_2020,Gherghescu_IveGot99_2024,Gherghescu_LookTrainingLarge_2024,Dubey_LlamaHerdModels_2024, OpenAI_GPT4TechnicalReport_2024}. To support hyper-scale \textit{Large Language Model (LLM)} training workloads, hyper-scale infrastructures are required~\cite{Duan_EfficientTrainingLarge_2024, Dubey_LlamaHerdModels_2024, Qian_AlibabaHPNData_2024,Zu_ResiliencyScaleManaging_2024,Jiang_MegaScaleScalingLarge_2024}; however, traditional network architecture is neither scalable nor cost-effective enough.

\begin{figure}[t]
  \centering
  \includegraphics[width=0.99\linewidth]{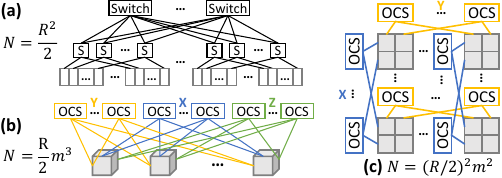}
  \caption{Scalability comparison of different topologies. $R$ is the switch radix, and $m$ is the local mesh/cube scale. (a) Two-level rail-optimized Fat-Tree; (b) OCS-based 3D-Torus; (c) RailX. \label{fig:overview}}
\end{figure}

Fat-tree-based topologies are widely used in existing datacenters; however, they are extremely expensive to provide sufficient bandwidth~\cite{Hoefler_HammingMeshNetworkTopology_2022,Barroso_DatacenterComputerDesigning_2019,Gherghescu_IveGot99_2024,Gherghescu_LookTrainingLarge_2024}, especially at hyper-scale. As shown in Figure~\ref{fig:overview}(a), the \textit{rail-optimized} network topology~\cite{Patronas_OpticalSwitchingData_2025, Qian_AlibabaHPNData_2024,Wang_RailonlyLowCostHighPerformance_2024,Liu_HostmeshMonitorDiagnose_2024,NVIDIA_NVIDIADGXSuperPOD_2023}, collaborating with high-bandwidth scale-up networks (\textit{e.g.}, \textit{NVLink} network~\cite{Ishii_NvlinkNetworkSwitchNvidias_2022,Tirumala_NVIDIABlackwellPlatform_2024}), reduces the number of spine/core switches while maintaining performance~\cite{NVIDIA_DoublingAll2allPerformance_2022}. However, the bandwidth and scalability are still limited by the high-radix packet switch and the scale-up network, both of which are expensive to further scale. Besides, multi-stage switching also introduces significant energy and latency overhead~\cite{Greenberg_CostCloudResearch_2008,Popoola_EnergyConsumptionSwitchcentric_2018,VictorAvelar_AIDisruptionChallenges_2023,Katebzadeh_EvaluationInfiniBandSwitch_2020}, potentially preventing system scaling.

Direct networks, though they do not require high-radix switches, also face the scalability challenge. The Torus topology naturally fits the traffic of traditional AI workloads~\cite{Jouppi_TPUV4Optically_2023, Google_TPUV4Document_2024} with data/tensor/pipeline parallelism (DP/TP/PP)~\cite{Shoeybi_MegatronLMTrainingMultiBillion_2020,Narayanan_EfficientLargescaleLanguage_2021,Jiang_MegaScaleScalingLarge_2024,Huang_GpipeEfficientTraining_2019}. However, with the parameter/activation size scaling, more parallel strategies, including sequence parallelism (SP)~\cite{Korthikanti_ReducingActivationRecomputation_2022}, expert parallelism (EP) for mixture-of-experts (MOE) models~\cite{Rajbhandari_DeepSpeedMoE_2022,Lepikhin_GShardScalingGiant_2020,Fedus_SwitchTransformersScaling_2022,Riquelme_ScalingVisionSparse_2021,Jiang_MixtralExperts_2024,DeepSeek-AI_DeepSeekV2StrongEconomical_2024,DeepSeek-AI_DeepSeekV3TechnicalReport_2024,Dai_DeepSeekMoEUltimateExpert_2024,DeepSeek-AI_DeepSeekR1IncentivizingReasoning_2025} and context parallelism (CP) for long sequences~\cite{Jacobs_DeepSpeedUlyssesSystem_2023,Liu_RingAttentionBlockwise_2023,NVIDIA_ContextParallelismOverview_2024}, are adopted and combined, making mapping on Torus complex. Moreover, the diameter of regular/twisted Torus networks is large, and the bisection bandwidth is insufficient for all-to-all traffic of MOE and ranking models~\cite{Camara_TwistedTorusTopologies_2010,Gangidi_RDMAEthernetDistributed_2024,Naumov_DeepLearningRecommendation_2019}. Furthermore, direct networks face flexibility and reliability challenges; thus, as shown in Figure~\ref{fig:overview}(b), \textit{Google TPUv4} cluster introduces optical circuit switches (OCSes) to reconfigure the interconnection~\cite{Zu_ResiliencyScaleManaging_2024}. However, the centralized optical switching layer limits the network scalability by the OCS port count (\textit{i.e.,} no more than 64 cubes by using 128-port OCSes~\cite{Liu_LightwaveFabricsAtScale_2023}).

There are a few other academic topologies, including \textit{HammingMesh}~\cite{Hoefler_HammingMeshNetworkTopology_2022}, \textit{BML}~\cite{Wang_ScalableHighPerformanceFaultTolerant_2020}, \textit{TopoOpt}~\cite{Wang_TopoOptCooptimizingNetwork_2023}, \textit{SiP-ML}~\cite{Khani_SiPMLHighbandwidthOptical_2021}, and \textit{Rail-only}~\cite{Wang_RailonlyLowCostHighPerformance_2024}, designed for AI training workloads.
However, they do not take the latest hyper-scale workloads with high-dimensional hybrid parallelism (especially expert parallelism) into account, and their scalability is still limited by the radix of packet/circuit switches (specifically, hard to scale to $>$100K chips with flat switching layer). Therefore, we are motivated to design a more scalable network architecture for modern hyper-scale LLM training workloads.


In this paper, we present the \textit{RailX} network architecture, where ``\textit{X}'' indicates the crossing of rails. As shown in Figure~\ref{fig:overview}(c), locally, chips within a node are directly interconnected by a high-bandwidth 2D-mesh, and the row/column rail ports of each node are connected to row/column circuit switches separately. By configuring the circuit switches, different row/column rails are interconnected into separate rings; simultaneously, nodes are interconnected into low-diameter topologies, including \textit{HyperX and Dragonfly}. The major contributions of this paper can be summarized as follows:
\begin{itemize}[leftmargin=15pt]
  \item \textit{RailX} fully utilizes advanced packaging/integration technologies and circuit switching, achieving ultra-high scalability and cost-effectiveness. Specifically, more than 100K chips can be interconnected with a flat (single-tier) 128-port circuit switching layer, and the typical diameter is only $2\sim 4$ inter-node hops. The network cost per injection/All-Reduce bandwidth of \textit{RailX} is less than $10\%$ of the Fat-Tree, and the cost per bisection/All-to-All bandwidth is less than $50\%$ of the Fat-Tree.
  \item We propose the \textit{Rail-Ring-based All-to-All} interconnection method based on \textit{Hamiltonian Decomposition} theory~\cite{Tillson_HamiltonianDecompositionK2m_1980} to construct an all-to-all topology from separate rings, achieving efficient All-Reduce and All-to-All communication simultaneously.
  \item We propose point-to-point routing and hierarchical collective algorithms to fully utilize local high-bandwidth and low-latency links, achieving better All-Reduce performance than traditional ring-collective algorithms on Torus and HammingMesh.
  \item We also introduce the \textit{Dimension Splitting} method to flexibly adjust the number of topology dimensions and the bandwidth/scale of each dimension, facilitating the mapping of diverse LLM training workloads with various types, shapes, scales, and parallelism (TP, CP, EP, DP, PP) strategies. In addition, \textit{RailX} can be used in MLaaS scenarios, where single or multiple training workloads can be flexibly mapped and scheduled, and failures can be worked around.
\end{itemize}

\section{Challenges and Motivations}
\subsection{Advances in Hardware Technologies}

In recent years, hardware technologies have made significant progress, promising to inspire new network architectures. Advanced integration technologies such as \textit{Panel-Level Packaging} and \textit{System-on-Wafer} allow many chips to be in integrated into a single package~\cite{Chun_InFO_SoWSystemonWaferHigh_2020,DouglasYu_TSMCPackagingTechnologies_2021,KevinZhang_OptionCoWoSSystemWafer_2024,Wang_DemonstrationWaferlevelIntegration_2023,Lau_RecentAdvancesTrends_2019}, and high-speed interfaces provide high-bandwidth, low-power, and low-latency connectivity between chips~\cite{Shukla_ShortReachInterconnect_2022,_UniversalChipletInterconnect_2024,Wei_93NVLinkC2CCoherent_2023,Tonietto_FutureShortReach_2022}. As shown in Table~\ref{tab:hop}, the \textit{UCIe} die-to-die interface provide $1317 GBps/mm$ shoreline bandwidth density ($1350 GBps/mm^2$ area density) within the package~\cite{_UniversalChipletInterconnect_2024}, much larger and cheaper than any traditional inter-chip interfaces. The latency and power consumption of such on-package interfaces are also much lower. At the same time, the \textit{Co-Packaged Optics (CPO)} technology~\cite{Minkenberg_CopackagedDatacenterOptics_2021,Maniotis_ExploringBenefitsUsing_2024} also significantly increases the off-package bandwidth density up to $128 GBps/mm$ (32 400G ports per chip edge)~\cite{Fathololoumi_TbOpticalCompute_2024, Mehta_AIComputeASIC_2024} and eliminates separate optical modules.

\begin{table}[ht]
  \fontsize{9pt}{10pt}\selectfont
  \centering
  \renewcommand\arraystretch{1.3}
  \begin{tabular}{|c|c|c|}
    \hline
    \textbf{Interface}         & \textbf{SerDes}    & \textbf{UCIe} \\
    \hline
    \textbf{Physical Medium}   & Fiber/Copper Cable & RDL/Substrate \\
    \hline
    \textbf{Density} (GBps/mm) & $128$              & $1317$        \\
    \hline
    \textbf{Latency} (ns)      & $100 + \text{ToF}$ & $2$           \\
    \hline
    \textbf{Energy} (pj/bit)   & $>5$               & $<1$          \\
    \hline
    \textbf{Channel Reach}     & $>10$ m            & $<25$ mm      \\
    \hline
  \end{tabular}
  \caption{Chip-to-chip interfaces~\cite{Mehta_AIComputeASIC_2024,Fathololoumi_TbOpticalCompute_2024,_UniversalChipletInterconnect_2024, Sella_FECKilledCutSwitch_2018, _CommonElectricalCEI_2022, Frankel_ProspectsOpticalTransceivers_2021, DavideTonietto_EnergyEfficiencySerial_2023} \label{tab:hop}}
\end{table}

However, there are also challenges in utilizing these technologies. Firstly, the wire density is high, and the connection distance is short; therefore, placement and wiring are challenging. Only flat topologies such as 2D-mesh and HexaMesh~\cite{Iff_HexaMeshScalingHundreds_2023} are feasible within the package~\cite{Pal_Designing2048Chiplet14336Core_2021, Hu_WaferScaleComputingAdvancements_2024,Chen_WaferscaleNetworkSwitches_2024}. At the same time, there is a significant gap between on-package and off-package bandwidth. As a result, when scaling out packages into large system-level networks~\cite{Talpes_MicroarchitectureDOJOTeslas_2023}, the mismatched bandwidth can lead to low utilization. Besides, mapping highly parallelized AI training workloads on such networks is also challenging.

\subsection{Challenges of Existing Networks for AI}
\begin{figure}[tb]
  \centering
  \includegraphics[width=0.99\linewidth]{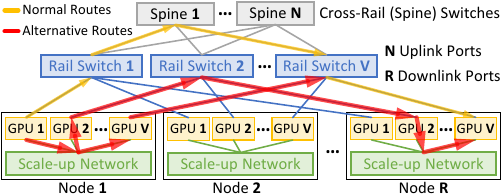}
  \caption{Rail-optimized Fat-Tree. The $i$-th GPU in each node is connected to the $i$-th rail switch. \label{fig:rail-optimized}}
\end{figure}
\subsubsection{\bf Fat-Tree is expensive}
In existing architectures, processors within a node are locally interconnected by a high-bandwidth low-latency Fat-Tree (\textit{e.g.,} \textit{NVLink} network)~\cite{Ishii_NvlinkNetworkSwitchNvidias_2022,Elster_NvidiaHopperGPU_2022,Choquette_NVIDIAHopperH100_2023}. 
As shown in Figure~\ref{fig:rail-optimized}, in a rail-optimized Fat-Tree, the NIC/GPU-$i$ of each node is connected to the same rail switch $i$ ($i=1,...,V$, where $V$ is the number of GPUs per node). With proper mapping, most communication does not need to go through the cross-rail switches; thus, the traffic imbalance and the cost of the network (number of cross-rail switches) are reduced~\cite{Wang_RailonlyLowCostHighPerformance_2024}. For massive cross-rail communication (\textit{e.g.,} all-to-all communication conducted by MOE and ranking models), local networks can be utilized to provide bandwidth among different rails~\cite{Yu_MoESysDistributedEfficient_2024,Rajbhandari_DeepSpeedMoE_2022, NVIDIA_DoublingAll2allPerformance_2022}. As the example shown in Figure~\ref{fig:rail-optimized}, messages from Node-1-GPU-1 can be sent to Node-1-GPU-V through the local network first, then to Node-R-GPU-V without going through the cross-rail switches.

Though the rail-optimized Fat-Tree is more cost-effective than the traditional Fat-Tree, the scalability challenge is still critical. Using rail switches with $R$ downlink ports and nodes with $V$ GPUs, a rail-optimized segment can only support $R\times V$ GPUs. The latest \textit{Alibaba HPN} achieves $400$ $Gb/s$ bandwidth among $15K$ GPUs by using $360$ $51.2T$ switches~\cite{Qian_AlibabaHPNData_2024}. At the same time, $7.5K$ \textit{NVLink} switches are required to provide $400\sim 900$ $GB/s$ local bandwidth. To further scale and increase the bandwidth, both the scale-out (Ethernet) and scale-up (NVLlink) networks are expensive. For example, on the one hand, the GB200 NVL72 node (cabinet) has an average sale price of \$$3$M while $36$ separate \textit{GB200 Superchips} have a total price of \$$2.16$M~{\cite{Say_NvidiaIncreasesBlackwell_2024}}; that is to say, the scale-up network accounts for about 30\% of the GPU node cost. On the other hand, to further interconnect the NVL72 node with 800G-per-GPU NIC in a 3-tier Fat-Tree, $72\times5\times800$G switching bandwidth and $72\times3$ 800G-links are required per node. If the prices of a $51.2$T switch and an optical cable with transceivers are estimated at \$$40$K~{\cite{NADDOD_64PortEthernetSwitch800Gb_2024}} and \$$2$K~{\cite{Gherghescu_IveGot99_2024,Gherghescu_LookTrainingLarge_2024,_NVIDIA800GbTwinPort_2025}}, the scale-out network introduces extra $21.9$\% cost. Besides, the energy ($>10$W$/Tbps/$hop~\cite{_MarvellTeralynx512T_2024}, additional one fifth of the computing power~\cite{Gherghescu_IveGot99_2024}) and latency ($>200ns/$hop~\cite{Katebzadeh_EvaluationInfiniBandSwitch_2020}) overhead of multi-layer switching are also significant. In summary, packet-switch-based Fat-Tree is too expensive to provide sufficient bandwidth at hyper scales.

  \begin{figure}[tb]
    \centering
    \includegraphics[width=0.95\linewidth]{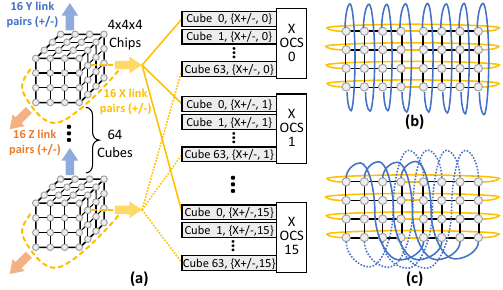}
    \caption{TPUv4 network architecture. (a) Each cube contributes two ports from opposite sides to each optical switch. (b) Regular Torus. (c) Twisted Torus. \label{fig:tpu}}
  \end{figure}
  \subsubsection{\bf Direct topologies have limited bisection bandwidth and flexibility} \label{sec:torus-limitation} AI training workloads with the three typical parallelism strategies (tensor, data, and pipeline) introduce 3D-shape communication patterns; therefore, Torus topology provides optimal ring-collective performance for traditional AI workloads without using expensive packet switches~\cite{Jouppi_TPUV4Optically_2023}. As shown in Figure~\ref{fig:tpu}, for usability and reliability purposes, the \textit{Google TPUv4} cluster introduces cheaper optical circuit switches to reconfigure interconnection among multiple 3D-cubes~\cite{Zu_ResiliencyScaleManaging_2024}. As a result, different shapes of regular/twisted 3D-Torus can be constructed for different workloads, and machine/chip/link failures can be worked around.

\begin{figure}[tb]
    \centering
    \includegraphics[width=0.9\linewidth]{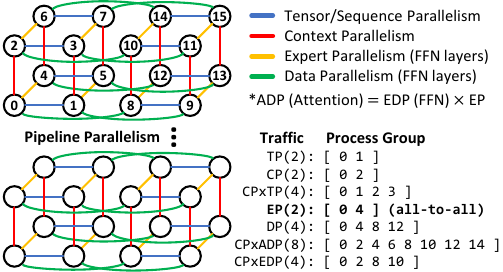}
    \caption{High-dimensional (TP, CP, EP, DP, PP) heterogeneous parallelism for hyper-scale LLM training. \label{fig:communication}}
  \end{figure}

  Though the Torus topology is cost-effective, it still faces scalability challenges. With the increase of model/data size, more parallel strategies, including expert parallelism with all-to-all traffic~\cite{Lepikhin_GShardScalingGiant_2020,Riquelme_ScalingVisionSparse_2021,Rajbhandari_DeepSpeedMoE_2022} and sequence/context parallelism~\cite{Liu_RingAttentionBlockwise_2023, Jacobs_DeepSpeedUlyssesSystem_2023,NVIDIA_ContextParallelismOverview_2024}, are adopted and combined~\cite{Dubey_LlamaHerdModels_2024,Zhu_LLaMAMoEBuildingMixtureExperts_2024,DeepSeek-AI_DeepSeekV3TechnicalReport_2024}. As shown in Figure~\ref{fig:communication}, a high-dimensional parallelism strategy is introduced for hyper-scale LLM training workloads. At the same time, attention and FFN (expert) layers adopt heterogeneous parallelism strategies due to the expert parallelism with all-to-all traffic (also see Figure~\ref{fig:expert-parallelism} in \S~\ref{sec:mapping})~\cite{Rajbhandari_DeepSpeedMoE_2022,Yu_MoESysDistributedEfficient_2024,DeepSeek-AI_DeepSeekV3TechnicalReport_2024}. The communication scope, volume, and frequency of different parallelisms are significantly different (\textit{e.g.}, TP usually introduces higher bandwidth requirement than DP~\cite{Dubey_LlamaHerdModels_2024, Wang_RailonlyLowCostHighPerformance_2024} and is hard to overlap with computation), but the dimension and bandwidth of the Tours topology is fixed and uniform. Therefore, mapping heterogeneous high-dimensional parallelism on Torus topologies is complex and inefficient.

  At the same time, the diameter of the Torus topology is large, and the bisection bandwidth is insufficient for all-to-all communication, which is essential for MOE and ranking models and also appears in context parallelism~\cite{Jacobs_DeepSpeedUlyssesSystem_2023}. As shown in Figure~\ref{fig:tpu}(c), to improve all-to-all performance, \textit{TPUv4} network adopts the twisted Torus, which crosswise connects rows/columns (over a $2n \times n$ mesh, connect $(2n$$-$$1,y)$ to $(0, y)$ and $(x, n$$-$$1)$ to $((x$$+$$n)\%2n, 0)$), thus improving bisection bandwidth by no more than two times~\cite{Camara_TwistedTorusTopologies_2010}. However, the twisted connection requires different cube rows/columns to be connected to the same OCSes, which constrains the network scale $N$ by the OCS radix $R$ ($N=R/2$) but only brings limited benefits ($< 2\times$ bisection bandwidth improvement). In addition, ring-collective on Torus cannot utilize the potential high-bandwidth benefits offered by intra-node direct connectivity.

  \begin{figure}[b]
    \centering
    \includegraphics[width=0.99\linewidth]{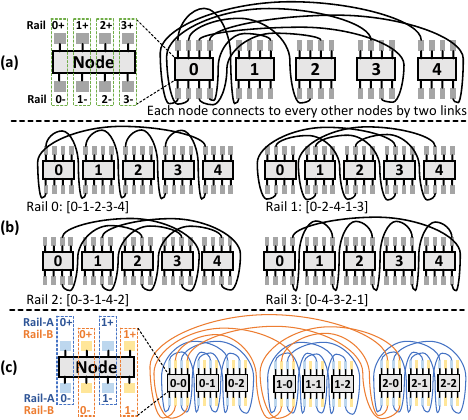}
    \caption{Rail-ring-based interconnection. (a) Each node is connected to every other node (b) $4$ rails are interconnected into rings separately with different orders. (c) Four rails are divided into two separate rail groups, forming a 2D-HyperX (2D all-to-all) topology with $3\times 3$ nodes. \label{fig:rail-interconnection}}
  \end{figure}

  \begin{figure*}[tbh]
    \centering
    \includegraphics[width=0.99\linewidth]{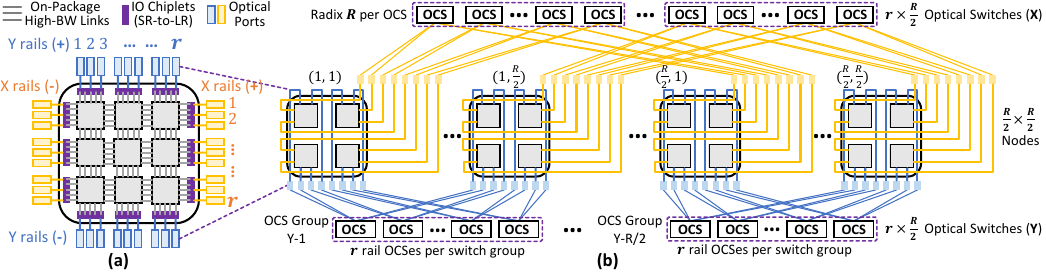}
    \caption{RailX physical architecture. (a) Chips within a node are connected into a 2D-mesh topology by high-bandwidth, low-latency direct links (\textit{e.g.,} UCIe~\cite{_UniversalChipletInterconnect_2024} and UALink~\cite{Synopsys_UALinkIPSolution_2024}). The short-reach interfaces at the edges are converted to long-reach optical ports for inter-node interconnection. (b) Nodes are connected to optical switches in a 2D organization ($\frac{R}{2}\times \frac{R}{2}$): different rows and columns of nodes are connected to different X/Y OCS groups ($r=mn$ OCSes per each group), and X/Y ports with the same rail-ID are connected to the same X/Y switch. \label{fig:architecture}}
  \end{figure*}
  \section{RailX Architecture}
  \subsection{Rail-Ring-based Interconnection}
  \label{sec:rail-ring}
  To address the challenges of traditional topologies, a new network architecture is motivated. One idea is inspired by the \textit{Hamiltonian Decomposition of Complete Graphs}~\cite{Tillson_HamiltonianDecompositionK2m_1980} that \textit{the edges of the complete directed graph of \,$k$ ($\neq4,6$) vertices can be partitioned into $k-1$ directed Hamiltonian cycles} (construction is shown in \S~\ref{appendix:construction}). As shown in Figure~\ref{fig:rail-interconnection}(a), a node has four rails (two $+/-$ ports per rail), and all inter-node links only connect the same rails. As shown in Figure~\ref{fig:rail-interconnection}(b), each rail forms a ring connecting all nodes in a different order. Consequently, five nodes form an all-to-all topology, and any two nodes are directly connected on two different rail rings (\textit{e.g.}, node-$0$ and node-$1$ are connected on rail-$0$ and rail-$3$). The formalized description of the \textit{rail-ring-based all-to-all interconnection} is given in Lemma~\ref{lemma:hamilton-decomp}.
  \begin{lemma}
    \label{lemma:hamilton-decomp}
    Given a node with $k-1$ rails (each rail has $+/-$ two ports), the all-to-all topology of \,$k$ ($\neq4,6$, only two exceptions) nodes can be constructed from $k-1$ rail-based Hamiltonian rings. Any two nodes $(A, B)$ are directly connected on two different rails $r_a$ and $r_b$: $A+\stackrel{r_a}{\Leftrightarrow}B-$ and $A-\stackrel{r_b}{\Leftrightarrow}B+$.
  \end{lemma}
  It is evident that such \textit{all-to-all} interconnection has a shorter diameter and better bisection bandwidth than Torus; however, the max all-to-all scale is limited by the number of rails (ports).

  Another idea is motivated by the \textit{Interface Grouping}~\cite{Feng_ScalableMethodologyDesigning_2023}, which splits ports into different groups to adjust the number of logical dimensions and the bandwidth per dimension. As shown in Figure~\ref{fig:rail-interconnection}(c), four rails can be divided into two separate rail groups, forming a 2D all-to-all (HyperX~\cite{Ahn_HyperXTopologyRouting_2009}) topology with $3\times 3$ nodes. A node with $n$ rails can be interconnected into a $k$-D hyper-X topology with $(\frac{n}{k}+1)^k$ nodes. It is also possible to let different rail groups have different numbers of rails and different interconnections. In this way, not only can the topology and bandwidth be flexibly adjusted, but the scalability is also improved.

  \subsection{Physical Architecture}
  \label{sec:topology}
  The following symbols are used in the description:
  \begin{table}[ht]
    \centering
    \renewcommand\arraystretch{1.1}
    \setlength{\tabcolsep}{2pt}
    \begin{tabular}{|c|p{7.2cm}|}
      \hline
      $m \times m$ & The scale of the 2D-mesh of chip in a node                      \\
      \hline
      $n$          & The number of off-package ports per chip edge                   \\
      \hline
      $r$          & The number of rails per dimension (X/Y), $r=mn$                 \\
      \hline
      $k$          & The multiple of on-package bandwidth over off-package bandwidth \\
      \hline
      $R$          & The radix (port number) of the optical circuit switch           \\
      \hline
      $N$          & The total number of chips                                       \\
      \hline
      $B_c$        & Bisection bandwidth (TX $+$ RX)                                 \\
      \hline
      $H_i, H_o$   & Internal direct hop and external optical hop                    \\
      \hline
    \end{tabular}
  \end{table}

  Basically, \textit{RailX} is a flat optical-circuit-switched network consisting of three physical levels: chip, node, and system. As shown in Figure~\ref{fig:architecture}(a), each chip is a processor/accelerator, and the original IO interfaces are high-density but short-reach (\textit{e.g.,} UCIe~\cite{_UniversalChipletInterconnect_2024}). $m\times m$ chips within a node are directly interconnected through these high-bandwidth links, forming a 2D-mesh topology. All the interfaces at the edges are converted by IO chiplets/modules to long-reach optical ports for upper-level interconnection~\cite{Chang_DOJOSuperComputeSystem_2022, Fathololoumi_TbOpticalCompute_2024, Mehta_AIComputeASIC_2024,Howard_FirstDirectMeshMesh_2023}. If a chip has $n$ ports at each edge, an entire node has $r=mn$ ports at each edge, and each pair of ports ($+/-$) for each row and column is called a rail. Advanced packaging technologies are optimal but not essential as long as the intra-node direct connectivity provides cheaper and higher bandwidth (\textit{e.g.,} NVLink and UALink~\cite{Synopsys_UALinkIPSolution_2024}) than inter-node long-distance links.

  As shown in Figure~\ref{fig:architecture}(b), all nodes in the system are connected to high-radix circuit switches in a 2D organization: X-rail $a$ and Y-rail $b$ of node $(i,j)$ are connected to X-OCS $(j, a)$ and Y-OCS $(i, b)$, respectively ($a,b\in[1,r], i,j\in[1, \frac{R}{2}]$). In other words, each node row and node column of the total $\frac{R}{2}\times \frac{R}{2}$ nodes are connected to one OCS group ($r$ switches), and ports with the same rail-ID are connected to the same switch.

  The physical architecture of \textit{RailX} is a circuit-switched \textit{HammingMesh}~\cite{Hoefler_HammingMeshNetworkTopology_2022} and is similar to \textit{TPUv4} but has two major differences. \textbf{1) \textit{Chips are locally connected into a higher-bandwidth 2D-mesh rather than a uniform-bandwidth mesh or cube.}} The short-reach direct connectivity can provide much higher and cheaper bandwidth with lower latency and energy consumption, which is not fully utilized by \textit{HammingMesh} and \textit{TPUv4}. Besides, the 2D-mesh with abundant bisection bandwidth is utilized as a high-bandwidth switch~{\cite{Feng_SwitchLessDragonflyWafers_2024,Chen_WaferscaleNetworkSwitches_2024}}; thus, low-diameter topologies can be constructed without introducing extra packet switches. In addition, the high-bandwidth 2D-mesh is regarded as a new hierarchy that facilitates more flexible mapping of training workloads with mixed and heterogeneous parallelism strategies, which will be further illustrated in \S~\ref{sec:high-dimension}, \ref{sec:collective-algorithm}, and \ref{sec:mapping}.

  \textbf{2) \textit{The nodes in RailX are connected to OCSes in a 2D organization rather than the centralized switching layer.}} One advantage of \textit{TPUv4} jointly connecting all cubes is to support the twisted Torus topology, which has a better bisection bandwidth. However, we have a better solution to provide a much higher bisection bandwidth; therefore, twisted Torus is not necessary so that we can interconnect different rows and columns separately to achieve high scalability. The total number of chips ($N$) and optical circuit switches ($N_s$) in \textit{RailX} is:
  \begin{equation}
    \label{eq:scale}
    \left\{
    \begin{array}{l}
      N = \left(\frac{R}{2}\right)^2m^2, \\
      N_s = rR.
    \end{array}
    \right.
  \end{equation}
  With existing hardware technologies, the scale of \textit{RailX} can be more than 100K chips ($R=128$~\cite{Liu_LightwaveFabricsAtScale_2023} and $m=5$~\cite{Talpes_MicroarchitectureDOJOTeslas_2023}, then $N=102400$). Even with normal PCB-level integration, the scale can still be more than 100K chips with higher-radix optical switches (\textit{e.g.}, $R=320$~\cite{Patronas_OpticalSwitchingData_2025,_PhotonicOpticalCircuit_2024} and $m=2$). In comparison, the \textit{TPUv4} pod (OCS-based 3D-Torus) can only support as many as $N_\text{TPUv4}=\frac{R}{2}m^3$ chips, limited by the 1D centralized switching layer.

  \subsection{Topology Configuration}
  Similar to \textit{TPUv4}, the circuit switches are configured at the beginning of a large training job, and on-chip routers (low-radix packet switches) are used for routing under each specific configuration. Therefore, the switching latency, which is the most significant disadvantage of optical circuit switches, is negligible. More fine-grained and dynamic reconfiguration (\textit{e.g.}, reconfiguration during each iteration) is also possible, which is briefly introduced in \S~\ref{sec:fine-grained-switching}. 
  
  By configuring optical switches, different logical topologies, including Torus, Draonglfy~\cite{Feng_SwitchLessDragonflyWafers_2024,Kim_TechnologyDrivenHighlyScalableDragonfly_2008}, and HyperX~\cite{Ahn_HyperXTopologyRouting_2009}, can be constructed and combined based on the interconnection method described in \S~\ref{sec:rail-ring}. The comparison of three base topologies is summarized in Table~\ref{tab:topology}.
  Topology configurations are according to the workload, trading off among scalability, diameter, and bisection bandwidth. Besides these three based topologies, other existing topology optimization methods based on optical switching, such as \textit{TopoOpt}~\cite{Wang_TopoOptCooptimizingNetwork_2023}, can also be utilized.

  \begin{table}[ht]
    \renewcommand\arraystretch{1.2}
    \begin{tabular}{cccc}
      \toprule
      \textbf{Topology} & \textbf{Torus}                 & \textbf{HyperX}  & \textbf{Dragonfly}        \\
      \midrule
      Scalability       & $\big( \frac{R}{2} \big)^2m^2$ & $(r$$+$$1)^2m^2$ & $(r$$+$$1)\frac{R}{2}m^2$ \\
      Diameter ($H_o$)  & $R$                            & $2 $             & $3$                       \\
      Bisection BW      & $\frac{16n}{Rm}$               & $\frac{2n}{m}$   & $\frac{2n}{m}$            \\
      \bottomrule
    \end{tabular}
    \caption{Comparison of different 2D topologies \label{tab:topology}}
  \end{table}

  \subsubsection{\bf 2D-Torus}
  The Torus topology can be naturally constructed by connecting the X-rails and Y-rails into parallel rings, just like the \textit{TPUv4}. Since Torus only connects adjacent nodes, all nodes in the system can be connected into an entire $\frac{R}{2}m\times \frac{R}{2}m$ 2D-Torus. Though Torus provides optimal all-reduce bandwidth, the diameter is up to $R\left(H_o+(m-1)H_i\right)$, and the theoretical upper-bound throughput~\cite{Dally_PrinciplesPracticesInterconnection_2004} for all-to-all communication is only
  \begin{equation}
    T_\text{Torus} = \frac{2B_c}{N} = \frac{4Rmn}{\left(\frac{R}{2}\right)^2m^2} = \frac{16n}{Rm}
  \end{equation}
  per chip, significantly decreasing with the scale. Therefore, 2D-Torus is not suitable for complex workloads with non-ring-collective traffic (\textit{e.g.}, MOE model with all-to-all traffic in the EP dimension).

  The workload mapping and collective algorithms of \textit{RailX-Torus} are different from the regular Torus, where each parallelism is aligned with dimension scales. Take the \textit{TPUv4} Pod slice with topology $8\times 16\times 16$ as an example, the tensor parallelism is 8-way mapped on the 8-scale dimension or 16-way mapped on a 16-scale dimension~\cite{Google_TPUV4Document_2024}. As for the \textit{RailX-Torus}, the high-bandwidth 2D-mesh can be utilized to apply hierarchical collectives~\cite{Cho_BlueConnectDecomposingAllreduce_2019} or as a high-bandwidth local dimension. Details will be illustrated in \S~\ref{sec:high-dimension} and \ref{sec:collective-algorithm}.

  \begin{figure}[tb]
    \centering
    \includegraphics[width=0.99\linewidth]{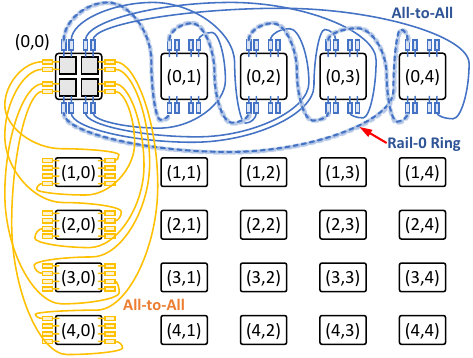}
    \caption{2D-HyperX configuration of 25 nodes. 4 X-rails and 4 Y-rails are separately configured into ring-based all-to-all (as shown in Figure~\ref{fig:rail-interconnection}).\label{fig:hyperx}}
  \end{figure}

  \subsubsection{\bf 2D-HyperX}
  The regular HyperX is a direct network of $S^D$ switches where switches belonging to the same dimension are fully connected~\cite{Ahn_HyperXTopologyRouting_2009}. For example, in an $S\times S$ HyperX, $S$ switches in each row and each column are all-to-all connected. Compared with the Torus, HyperX has a much shorter diameter (equals to the dimension number $D$) and a much higher bisection bandwidth.

  Based on the interconnection method described in \S~\ref{sec:rail-ring}, ``switch-less'' HyperX can be constructed in the \textit{RailX}. As shown in Figure~\ref{fig:hyperx}, the X-rails and Y-rails are separately interconnected into rail-ring-based all-to-all (two direct links between every node pair in every dimension), forming a $(r+1) \times (r+1)$ 2D-HyperX. Different from the regular HyperX, there are no extra high-radix packet switches in the \textit{RailX-HyperX}, and there are two direct links between every node pair in every dimension.

  The maximum scale of the RailX-2D-HyperX topology is $(r+1)^2m^2$, limited by the rail number per node. If $r$ is large enough (\textit{i.e.}, $r+1=\frac{R}{2}$), the entire system can be configured into a single 2D-HyperX. With advanced CPO technologies (\textit{e.g.}, Broadcom achieves 32 optical ports per chip edge~\cite{Mehta_AIComputeASIC_2024}), the rail number of a node can be even larger than the optical switch; thus there can be more physical channels between all-to-all pairs. All-to-all interconnection of $\frac{r}{a}+1$ nodes can be constructed by connecting every node pair with $2a$ rails.

  For the maximum size, the bisection bandwidth for each HyperX row/column is up to $4\left(\frac{r+1}{2}\right)^2$. Thus, the bisection throughput for all-to-all communication is
  \begin{equation}
    T_\text{HyperX}  = \frac{2\times (r+1)\times 4\left(\frac{r+1}{2}\right)^2}{(r+1)^2m^2} \approx \frac{2n}{m}
  \end{equation}
  per chip, more scalable than the 2D-Torus. The evaluation results are shown in \S~\ref{sec:a2a-performance}. Besides, diameter of the 2D-HyperX is only $2H_o+(5m-6)H_i$ (details are shown in \S~\ref{sec:p2p-routing}), also much smaller than 2D-Torus.

  \begin{figure}[tb]
    \centering
    \includegraphics[width=0.99\linewidth]{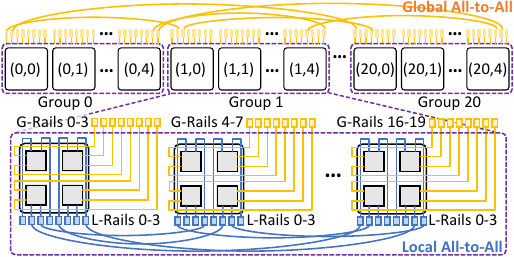}
    \caption{Dragonfly configuration. Five nodes are locally all-to-all interconnected by four local rails, forming a group with 20 global rails; then, 21 groups are globally all-to-all interconnected. \label{fig:dragonfly}}
  \end{figure}
  \subsubsection{\bf Dragonfly}
  The regular Dragonfly is another low-diameter topology where switches are locally fully connected within groups, and groups are also globally all-to-all interconnected. As shown in Figure~{\ref{fig:dragonfly}}, the local all-to-all connection of Dragonfly is the same as the first dimension of 2D-HyperX. However, for the second dimension (global interconnection), Dragonfly only connects one switch rather than all switches in each group to other groups. As a result, Dragonfly can support a much larger scale in case of very large circuit switch radix $R$ or very small rail number $r$.

  A local all-to-all group, consisting of $r+1$ nodes, has $r(r+1)$ rails for global interconnection. Therefore, the total number of groups is $\min(r^2+r+1, R/2)$. Consequently, it is possible to use two different kinds of optical switches: fast low-radix switches (\textit{e.g.}, AWGR)\cite{Fu_FirstDemonstrationMonolithic_2021,Mellette_RealizingRotorNetPractical_2024,Clark_SynchronousSubnanosecondClock_2020} and slow high-radix switches~\cite{_PhotonicOpticalCircuit_2024, Liu_LightwaveFabricsAtScale_2023}. This paper mainly discusses slow switching because it is more mature and practical. Specifically, if the local (Y) circuit switch radix is $R_l=2(r+1)$ and the global (X) circuit switch radix is $R_g=2[r(r+1)+1]$, the entire system can be configured into one single Dragonfly with $(r+1)(r^2+r+1)m^2$ chips. For the maximum size, the bisection throughput for all-to-all communication is
  \begin{equation}
    T_\text{Dragonfly}  = \frac{2\times (r+1)\frac{R}{2} \times r}{(r+1)\frac{R}{2}m^2} \approx \frac{2n}{m}
  \end{equation}
  per chip, the same as 2D-HyperX, also much scalable than 2D-Torus.

  \begin{figure}[b]
    \centering
    \includegraphics[width=0.99\linewidth]{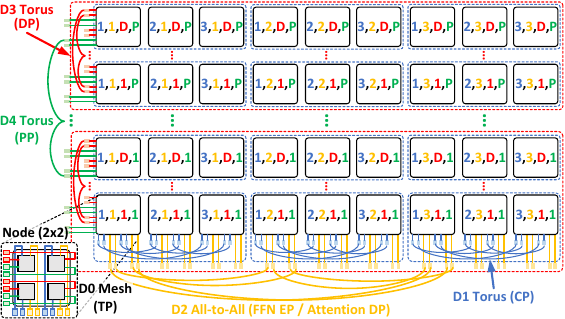}
    \caption{High-dimensional heterogeneous topology ($4\times 3 \times 3 \times D \times P$) and the mapping of the training workload ($TP \times CP \times EP \times DP \times PP$). A more detailed version is shown in \S~\ref{appendix:example} Figure~\ref{fig:example}. \label{fig:high-dimension}}
  \end{figure}
  \subsubsection{\bf High-dimensional heterogeneous topology} \label{sec:high-dimension}
  As mentioned in \S~\ref{sec:torus-limitation}, high-dimensional heterogeneous parallelism is introduced for hyper-scale LLM training workloads; thus, the 2D topology is insufficient. Based on the second interconnection method described in \S~\ref{sec:rail-ring}, a high-dimensional topology can be constructed by splitting the rails of the two physical dimensions into multiple logical dimensions. As the example shown in Figure~{\ref{fig:high-dimension}}, the $r$ rails of each dimension are split into two groups with $\frac{r}{2}$ rails, and each rail group can be interconnected into Torus (unlimited scale) or rail-ring-based all-to-all (maximum scale $\frac{r}{2}+1$). As a result, by using $r=4$ nodes, a 5D network with $4\times 3 \times 3 \times D \times P$ chips can be constructed, where the $0$-th dimension is the 2D-mesh within the node, the $2$-rd dimension is all-to-all with scale $3$, and $1/3/4$-th dimension is Torus with scale $3/D/P$. The total scale of two split dimensions is still limited by the optical switch radix: $D\times P \leq \frac{R}{2}$. For balance purpose, we typically split the $n$ rails of each chip row/column with the same proportion; thus, all chips have the same direct bandwidth in each dimension. Alternatively, we can also take the entire node as a whole and unevenly split the rails. For example, we can split the $2\times 2$ rails in Figure~\ref{fig:high-dimension} into a $3:1$ proportion.

  According to the shape of workloads (number of communication dimensions and the traffic volume of each dimension), we can flexibly configure the topology and adjust the bandwidth of each dimension. For example, if the traffic volume of two dimensions are $V_1$ and $V_2$ respectively, we can adjust the bandwidth by splitting the $n$ rails into $n_1$ and $n_2$ ports per rail group and minimize the total communication time $T_{Total}=\frac{V_1}{n_1} + \frac{V_2}{n_2}$ or the slowest time $T_{Slow}=\max(\frac{V_1}{n_1}, \frac{V_2}{n_2})$. If the communication of two dimensions is non-overlapped, we can even allocate all the bandwidth to one dimension first and then to the other dimension by reconfiguring the optical switches. More details will be discussed in \S~{\ref{sec:mapping}}.



  \subsubsection{\bf 2D-mesh as high-bandwidth virtual switch} \label{sec:high-bandwidth-mesh} All low-diameter topologies illustrated above are constructed by utilizing the 2D-mesh as high-radix switches~\cite{Chen_WaferscaleNetworkSwitches_2024, Feng_SwitchLessDragonflyWafers_2024}. As a result, extra intra-node traffic is introduced for inter-node communication. To prevent the 2D-mesh from becoming the bottleneck, the intra-node bandwidth is supposed to be larger. Estimated by the bisection bandwidth per port
  \begin{equation}
    \frac{2\times kmn}{4mn} > 1 \Rightarrow k > 2,
  \end{equation}
  the intra-node bandwidth of the 2D-mesh should be at least twice the inter-node bandwidth. According to our evaluations in \S~\ref{sec:a2a-performance}, $2\sim 4\times$ higher bandwidth is sufficient to provide non-blocking switching capability, which is easy to achieve by using short-reach direct links.





  \section{Communication Algorithms}

  Due to the good features of rail-ring-based all-to-all interconnection, many existing algorithms can be borrowed from Torus, Dragonfly, HyperX, and HammingMesh topologies. The major difference is the high-bandwidth local 2D-mesh, which brings benefits but also introduces routing challenges.

  \subsection{Point-to-Point Routing}
  \label{sec:p2p-routing}

  \subsubsection{\bf Minimal-Routing} 
\begin{figure}[tb]
    \centering
    \includegraphics[width=0.999\linewidth]{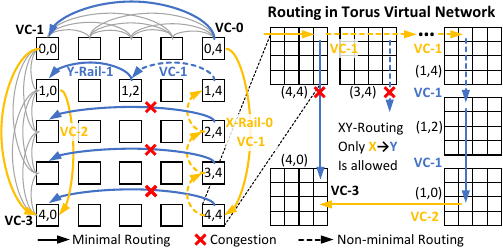}
    \caption{Minimal and non-minimal adaptive routing from $(0,4)$ to $(4,0)$ on 2D-HyperX. \label{fig:routing}}
  \end{figure}

Minimal routing algorithms on RailX-based HyperX are similar to traditional HyperX~\cite{Ahn_HyperXTopologyRouting_2009, McDonald_PracticalEfficientIncremental_2019}, but extra routing hops are required over the local 2D-mesh. Two minimal routes from source node $(0,4)$ to destination node $(4,0)$ are drawn in Figure~\ref{fig:routing}: messages traverse to node $(0,0)$ or $(4,4)$ first and then to node $(4,0)$. Since nodes are connected by two links on both mesh sides (see Figure~\ref{fig:hyperx}), if we always choose the nearest inter-node link when routing on 2D-mesh, a packet travels at most $(\frac{m}{2}-1)+(m-1)$ internal hops at each non-destination node and at most $2(m-1)$ internal hops at the destination node. Therefore, the diameter of the 2D-HyperX is no more than $2H_o+(5m-6)H_i$. The on-mesh routing algorithm can be deterministic or adaptive (\textit{e.g.}, dimension-order or north-last~\cite{Glass_TurnModelAdaptive_1992}) based on one single virtual channel (VC). As long as we increase VC at each node hop along the routes, any deterministic or adaptive minimal routing algorithm on HyperX is deadlock-free. A specific example is given in Algorithm~\ref{alg:deterministic}, where $(X,Y)$ and $(x,y)$ are the node coordinate in 2D-HyperX and the chip coordinate in 2D-mesh, respectively. For other topology configurations, similar methods can be applied, and the required VC number for deadlock-free minimal routing equals the inter-node diameter ($d_o$) plus one.

  \begin{algorithm}[tb]
    \begin{algorithmic}[1]
      \REQUIRE Current chip coordinate: $(X_c, Y_c, x_c, y_c)$;\\
      \quad Destination chip coordinate: $(X_d, Y_d, x_d, y_d)$.
      \IF[Arriving Destination]{$X_c = X_d$ \& $Y_c = Y_d$}
      \STATE {\scshape Routing-on-Mesh$(x_c, y_c, x_d, y_d)$} by VC-2
      \ELSIF[X-Rail-First]{$X_c \neq X_d$}
      \STATE $(x', y') \leftarrow$ the nearest chip links to $(X_d, Y_c)$
      \IF{$(x_c, y_c)=(x',y')$}
      \STATE {\scshape X-Rail$(X_c, X_d)$} by VC-1
      \ELSIF{$(x_c, y_c)\neq(x',y')$}
      \STATE {\scshape Routing-on-Mesh$(x_c, y_c, x', y')$} by VC-0
      \ENDIF
      \ELSIF{$X_c = X_d$ \& $Y_c \neq Y_d$}
      \STATE $(x'', y'') \leftarrow$ the nearest chip links to $(X_d, Y_d)$
      \IF{$(x_c, y_c)=(x'',y'')$}
      \STATE {\scshape Y-Rail$(Y_c, Y_d)$} by VC-2
      \ELSIF{$(x_c, y_c)\neq(x'',y'')$}
      \STATE {\scshape Routing-on-Mesh$(x_c, y_c, x'', y'')$} by VC-1
      \ENDIF
      \ENDIF
    \end{algorithmic}
    \caption{\scshape Deterministic Minimal Routing \label{alg:deterministic}}
  \end{algorithm}

  \subsubsection{\bf Non-minimal Adaptive Routing} Minimal routing is not always optimal when traffic is unbalanced (\textit{e.g.}, all-to-all traffic of MOE models~\cite{Chen_UnderstandingMixtureexpertsLayer_2022}); therefore, non-minimal adaptive routing algorithms are necessary~\cite{McDonald_PracticalEfficientIncremental_2019,Kim_TechnologyDrivenHighlyScalableDragonfly_2008}. For any routing algorithm, we can always adopt the same method as the minimal routing: increasing VC at each node hop along the routes. However, the required VC number is as many as the length of the longest route.

  Due to the good features of rail-ring-based interconnection (2D-rings), we can leverage the Torus routing algorithm to achieve deadlock-free non-minimal routing on RailX with fewer VCs. Previous studies have shown that only two VCs are required for deadlock-free XY routing on 2D-Torus~\cite{Dally_PrinciplesPracticesInterconnection_2004}, and only one single VC is sufficient if using \textit{Bubble} flow-control techniques~\cite{Puente_AdaptiveBubbleRouter_2001,LizhongChen_WormBubbleFlowControl_2013}. The basic idea is to fully utilize each VC by using XY-routing-based Torus virtual networks where packets are routed along the X-rail and then the Y-rail. As shown in Figure~\ref{fig:routing}, packets use VC-0 at the source node $(0,4)$ and then increase to VC-1 at the intermediate node $(4,4)$. If the direct channel between $(4,4)$ and $(4,0)$ is congested, adaptive misrouting can be performed. As long as the packet follows the XY-routing algorithm on Torus, it can be freely routed in the VC-1 network. In the figure, the packet is routed to node $(1,4)$ along the X-rail-0 and then routed to node $(1,0)$ along the Y-rail-1. Once the packet wants to change to X-rails, which violates the XY-routing algorithm, it must increase to VC-2. Finally, it is routed to the destination at node $(4,0)$ through VC-3. With this method, we combine limited times ($a>d_o$) of free routing and infinite times of Torus routing to achieve deadlock-free non-minimal routing on RailX with a limited number ($a+1$) of VCs.


  \subsection{Collective Communication Algorithm}
  \label{sec:collective-algorithm}
  For AI training workloads, \textit{All-Reduce} algorithms, which equals to one \textit{Reduce-Scatter} followed by one \textit{All-Gather}, are essential. Due to the rail-ring-based interconnection, existing collective algorithms on Torus such as \textit{Ring}~\cite{AndrewGibiansky_BringingHPCTechniques_2017} and \textit{Swing}~{\cite{Sensi_SwingShortcuttingRings_2024}} can be directly applied to RailX topology. The communication time of the bidirectional-ring-based Reduce-scatter and All-gather algorithm can be estimated as~\cite{Cho_BlueConnectDecomposingAllreduce_2019}
  \begin{equation}
    T_R (p, V, B) = (p-1)\alpha + \frac{p-1}{p}\frac{V}{2B},
  \end{equation}
  where $p$ is the number of processors, $\alpha$ is the step (hop) latency, $V$ is the data volume, and $B$ is the bandwidth.

  The 2D-ring-based algorithm on the RailX network splits data into two chunks and simultaneously executes the hierarchical algorithm~\cite{Cho_BlueConnectDecomposingAllreduce_2019} in both X and Y dimensions~\cite{Hoefler_HammingMeshNetworkTopology_2022, Rashidi_ThemisNetworkBandwidthaware_2022}. Therefore, the communication time of the 2D-ring-based All-Reduce algorithm on the $m^2 \times p\times p$ RailX (2D-Torus or 2D-HyperX) can be estimated as \footnote{$\approx$: we omit lower order term $o(p)$ and on-package hop latency}
  \begin{equation}
    \begin{aligned} T_\text{2D-Ring} & = 2\left[T_R\left(mp, \frac{V}{2}, nB\right)+T_R\left(mp, \frac{V}{2mp}, nB\right)\right] \\[-2pt]
                                 & \approx 4mp\alpha + \frac{V}{2nB},
    \end{aligned}
  \end{equation}
  where $\alpha$ is the inter-node optical hop latency, $nB$ is the total bandwidth per chip edge. Though many of the hops along the ring are on-package hops, the total communication time is limited by the slowest hop, \textit{i.e.}, the inter-node hop.

  \begin{figure}[tb]
    \centering
    \includegraphics[width=0.9\linewidth]{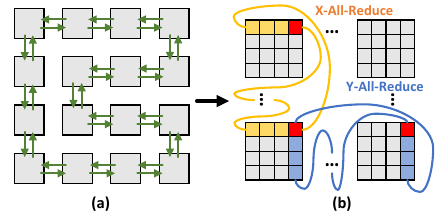}
    \caption{Hierarchical All-Reduce algorithm on RailX. (a) Bidirectional-ring-based All-Reduce on local 2D-mesh. (b) 2D-ring-based All-Reduce.  \label{fig:collective}}
  \end{figure}

  \label{sec:hierarchical-allreduce}
  To fully leverage the high on-package bandwidth in RailX, we present a hierarchical All-Reduce algorithm. As shown in Figure~\ref{fig:collective}(a), we can first perform local All-Reduce algorithms~\cite{Luczynski_OptimalWaferScaleReduce_2024,Kumar_HighlyAvailableData_2020} (\textit{e.g.}, bidirectional ring\cite{Laskar_EnhancingCollectiveCommunication_2024}) over the high-bandwidth 2D-mesh. Then, as shown in Figure~\ref{fig:collective}(b), global All-Reduce is performed. All the chips with the same local rank ID are still connected by the same X/Y-rail, and the inter-node bandwidth is shared by $m$ chips (local ranks) along the rail. The communication time of the hierarchical All-Reduce algorithm on the $m^2 \times p \times p$ RailX network is estimated as
  \begin{equation}
    \begin{aligned}
      T_\text{RailX} & \approx 2\times\frac{V}{2\boldsymbol{k}nB} + \left(4p\alpha + \frac{V/\boldsymbol{m^2}}{2nB/\boldsymbol{m}} \right) \\[-2pt]
                     & = 4p\alpha + \left(\frac{2}{k}+\frac{1}{m}\right)\frac{V}{2nB},
    \end{aligned}
  \end{equation}
  where $k$ is the multiple of on-package bandwidth (in local 2D-mesh) over off-package bandwidth. We can see that the on-package bandwidth is utilized, and only $p$ rather than $mp$ inter-node steps are required. Typically, as long as $k>2$, our hierarchical algorithm on RailX achieves better global All-Reduce performance than the traditional 2D-ring-based algorithms. The evaluation results are shown in \S~\ref{sec:all-reduce-performane}. We can also utilize the all-to-all interconnection in HyperX configuration to achieve a lower-latency All-Reduce algorithm, whose latency does not increase with the scale $p$. Limited by the space, it is shown in \S~\ref{appendix:a2a-allreduce}.

  As for specific mapping of training workloads, if the tensor parallelism is mapped on the high-bandwidth 2D-mesh, which is the typical mapping, the communication of other parallelisms (\textit{e.g.}, data parallelism) is among chips in different nodes with same local rank ID, and the inter-node bandwidth is shared by $m$ chips. Therefore, the communication time of 1D (among $p$ nodes) and 2D (among $p\times p$ nodes) node-level All-Reduce algorithm is estimated as

  \begin{equation}
    T_{1D} \approx 2p\alpha + \frac{V}{nB/m},\; T_{2D} \approx 4p\alpha + \frac{V}{2nB/m}.
  \end{equation}
  As mentioned in \S~\ref{sec:high-dimension}, we split the rails of the two physical dimensions to construct high-dimensional logical topologies. Collective communication may be performed among multiple dimensions. For example, the communication of QKV layers (non-expert layers) data parallelism is among CP/EP/DP dimensions (see \S~\ref{sec:mapping} and \S~\ref{appendix:example}~Table~\ref{tab:communication}). Similar methods can be adopted for high-dimensional All-Reduce over high-dimensional topologies, and we denote the communication time by $T_{h\text{D}}(n_1, ..., n_{h})$, where $n_i\times B$ is the bandwidth of dimension $i$. Besides our algorithms, there are other fine-grained collective scheduling algorithms~\cite{Shah_TACCLGuidingCollective_2023, Liu_RethinkingMachineLearning_2024,Wu_MCCSServicebasedApproach_2024, Won_TACOSTopologyAwareCollective_2024, Rashidi_ThemisNetworkBandwidthaware_2022, Rajasekaran_CASSININetworkAwareJob_2024} that can be applied.

  \section{Mapping and Scheduling}
  \label{sec:mapping}

  \begin{figure}[t]
    \centering
    \includegraphics[width=0.9\linewidth]{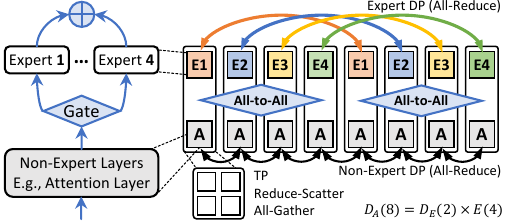}
    \caption{Expert parallelism. \label{fig:expert-parallelism}}
  \end{figure}

  4D parallelism (TP, CP, PP, DP) is used to train the latest Llama 3 405B model~\cite{Dubey_LlamaHerdModels_2024}. On this basis, we consider a larger-scale model with the MOE structure and introduce the expert parallelism (EP) over the expert (FFN) layers~\cite{Rajbhandari_DeepSpeedMoE_2022,Zhu_LLaMAMoEBuildingMixtureExperts_2024,Ishii_NvlinkNetworkSwitchNvidias_2022,Singh_HybridTensorExpertDataParallelism_2023,DeepSeek-AI_DeepSeekV3TechnicalReport_2024}. For simplicity's sake, we assume both the attention and FFN layers adopt the same TP, CP, and PP scale. Therefore, as shown in Figure~{\ref{fig:expert-parallelism}}, the DP scale of attention layers equals the total scale of DP and EP of FFN layers ($D_a=ED_e$). The order of parallelism dimensions is $[T, C, E, D_e, P]$, where the innermost tensor parallelism usually introduces the most massive communication traffic. One mapping example is illustrated in \S~\ref{sec:high-dimension}, and the communication pattern and time of each parallelism are shown in \S~\ref{appendix:example} Table~\ref{tab:communication}. Many existing scheduling methods, such as \textit{CASSINI}~\cite{Rajasekaran_CASSININetworkAwareJob_2024}, can be borrowed for mapping and scheduling on \textit{RailX}. Limited by the space, this paper only focuses on the bandwidth allocation scheduling for the \textit{Dimension Spliting} method.

  \subsection{Static Allocation}
  \label{sec:bandwidth-allocation}
  As discussed in \S~\ref{sec:high-dimension}, we split the rails of each node into multiple dimensions according to the shape of workloads. The static allocation is to configure the network topology at the beginning of a training job and not reconfigure within training iterations, which is a common way with ``slow-switching'' OCSes~\cite{Liu_LightwaveFabricsAtScale_2023, Wang_TopoOptCooptimizingNetwork_2023}.

  In practice, the factors affecting real efficiency are very complex. Besides the communication efficiency, the computation efficiency and the communication-computation overlap are also important. A highly generalized expression
  \begin{equation}
    T_\text{Actual} \approx T_\text{Comp} + \max\left\{T_\text{Comp}^*, T_\text{Comm}^*\right\} + T_\text{Comm}
  \end{equation}
  is used to estimate the total time, where $T_\text{Comp}$/$T_\text{Comm}$ is the computation/communication time that cannot be overlapped, $T_\text{Comp}^*$/$T_\text{Comm}^*$ is the time that can be overlapped. As the example shown in Figure~\ref{fig:high-dimension}, if we split the bandwidth for DP and PP ($n=n_d+n_p$), both of which can be overlapped, the actual total time we want to minimize is
  \begin{equation}
    \text{arg}\min_{n_d,n_p} \left\{\max\left(T_{\text{Comp},d}^*,\frac{V_d}{2n_dB}\right)+\max\left(T_{\text{Comp},p}^*,\frac{V_p}{2n_PB}\right)\right\},
  \end{equation}
  where, $V$ is the data volume of each parallelism, and $B$ is the bandwidth per each port.

  During the entire training process, the mapping strategies and bandwidth allocation can still be adjusted. For example, in Llama 3 405B training, the context length gradually increases from 8K to 128K in six stages~\cite{Dubey_LlamaHerdModels_2024}, and the context parallelism is introduced only for long context length. Therefore, the bandwidth allocation may also be adjusted when the parallelism and mapping change. However, such adjustments are not frequent, so they are still regarded as static allocations. Exploration results are shown in \S~\ref{sec:dimension-splitting-results}.

  \subsection{\bf Dynamic Allocation}
  \label{sec:fine-grained-switching} Based on static allocation, which splits the bandwidth of one physical dimension into two logical dimensions in a fixed way, we tend to map the highest and lowest communication to one dimension. \textit{E.g.}, if the communication requirement is $EP>CP>DP>PP$, the optimal allocation is to map the EP and PP to one physical dimension and the CP and DP to the other physical dimension.

  \begin{figure}[b]
    \centering
    \includegraphics[width=0.95\linewidth]{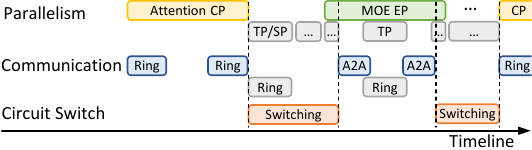}
    \caption{Circuit switching to allocate all bandwidth between context parallelism and expert parallelism. \label{fig:cp-ep}}
  \end{figure}

  However, if the two communications are separated by other communication or computation, we can use circuit switching to allocate the bandwidth dynamically. As shown in Figure~\ref{fig:cp-ep}  (also shown in \S~\ref{appendix:trace}~Figure~\ref{fig:trace}), CP and EP communications are non-overlapped. They are separated by the computation of attention input/output projection, the communication of SP/TP, and the pre/post-process of MOE layers~\cite{_NVIDIAMegatronLM_2025,NVIDIA_ContextParallelismOverview_2024}. The time interval is about a few milliseconds for the small-scale model and can be longer for larger models. Therefore, it is possible to dynamically configure the circuit switch and allocate the bandwidth for CP and EP communications. As a result, both CP and EP communications can fully utilize the bandwidth of one entire physical dimension.

  \section{Evaluation}

  \subsection{Methodology}

  We use a cycle-based network simulator and a hardware-validated analytical model to evaluate \textit{RailX}.

  \subsubsection{\bf Analytical Model and Trace Synthesis} Due to the impracticality of collecting real traces from extreme-scale AI training workloads, we build a hardware-validated analytical model to give estimates of the computation and communication. To validate the analytical model, we run training workloads with small-scale hybrid parallelism on real hardware and collect the trace by using NVIDIA Nsight Systems~\cite{NVIDIA_NVIDIANsightSystems_2024}. Details are shown in \S~\ref{appendix:trace}. The computation time of larger tiles/blocks is estimated based on the computing FLOPS (multiples of FLOPS $\times$ measured time of small tiles/blocks). The volume and scopes of major communication are directly extracted from the training framework and validated by the Nsight traces.

  \subsubsection{\bf Cycle-based Simulator} We build a cycle-based network simulator based on the  CNSim framework~\cite{Feng_EvaluatingChipletbasedLargeScale_2024}, which supports fine-grained microarchitecture modeling and multithreading (also see \S~\ref{appendix:cnsim}), to evaluate the communication performance of \textit{RailX}. To simplify and accelerate the simulation, we omit complex network protocols and use the ideal virtual cut-through router and the credit-based lossless flow control~\cite{Dally_PrinciplesPracticesInterconnection_2004}. The bandwidth of the slowest link (inter-node link) in the network is normalized to 1 flit/cycle (\textit{e.g.,} 64 Gb/s bandwidth at 8B flit size and 1GHz frequency), and the on-chip/intra-node link bandwidth is $k$ multiples. The default input buffer size per virtual channel of the router is 16 flits (\textit{i.e.,} maximum message size). The evaluated traffic patterns are synthesized by the hardware-validated analytical model. The inter-node latency is set to 10 cycles, while the intra-node latency is set to 1 cycle. 

  \subsection{Cost Analysis}\
  \label{sec:cost-analysis}
  We compare the cost of several typical systems: 1) non-blocking/tapered Fat-Tree; 2) HammingMesh (Hx$a$Mesh with $a$$\times$$a$ boards) ; 3) 3D-Torus without OCS; 4) TPUv4 (OCS-based 3D-Torus); 5) 2D Fat-Tree (Rail-Only~\cite{Wang_RailonlyLowCostHighPerformance_2024}, see \S~\ref{appendix:related-work}); 6) RailX$a$Mesh. For a fair comparison, we assume all chips have 1.8TB/s (36$\times$400G) off-package bandwidth; specifically, 36 rails for Fat-Tree, 18 rails for 2D Fat-Tree, 9 planes for HammingMesh, 6 ports per direction for 3D-Torus, and $n=9$ rails per chip for RailX. Short-reach package/PCB-level connectivity is used for 2D-mesh; passive copper cables (PCCs) are used for 3D-cube (inter $2\times2$ mesh boards~\cite{Jouppi_TPUV4Optically_2023}); and active optical cables are used for inter-mesh/cube/switch connections. The cost is estimated based on the components, and we assume OCS has twice as many ports as electrical packet switches at the same cost~\cite{Patronas_OpticalSwitchingData_2025,Poutievski_JupiterEvolvingTransforming_2022,Liu_LightwaveFabricsAtScale_2023,Wang_TopoOptCooptimizingNetwork_2023,Urata_MissionApolloLanding_2022}. A passive 400G copper cables costs \$250~\cite{_NVIDIAPassiveCopper_2025}, an 400G active optical transceiver (AOT) costs \$1000~\cite{_NVIDIA800GbTwinPort_2025,Gherghescu_IveGot99_2024,Gherghescu_LookTrainingLarge_2024}, a 64-port 400G packet switch or a 128-port OCS costs \$35K~\cite{_MellanoxQuantum2QM9700_2025}. The cost of short-reach package/PCB-level connectivity is neglected (included in chips). It should be noted that, with $n=9$, $m=7$ results in $r=63$ to match the OCS radix $R/2=64$.

\begin{table}[tb]
  \fontsize{9pt}{10pt}\selectfont
  \centering
  \setlength{\tabcolsep}{2pt}
  \caption{Scalability/cost comparison (packet/circuit switch radix is 64/128, details are in \S~\ref{appendix:cost}~Table~\ref{appendix:tab:cost}). \label{tab:cost}}
  \begin{tabular}{lccccc}
    \toprule
    \textbf{Topology} & \textbf{Scale}  & \textbf{Cost}  & \textbf{Cost}                      & \textbf{Glob. BW}    & \textbf{Cost}                      \\
                      & \textbf{[\#]}   & \textbf{[M\$]} & \textbf{[\,/\,Inject]}             & \textbf{[\% Inject]} & \textbf{[\,/\,GBW]}                \\
    \midrule
    2-Tier FT         & 2048            & 415.9          & 1$\times$                          & 100                  & 1$\times$                          \\
    1:3 Tap. FT       & 3072            & 395.7          & 0.65$\times$                       & 33.3                 & 1.90$\times$                       \\
    1-FT Hx4Mesh      & 16384           & 375.6          & 0.11$\times$                       & 12.5                 & 0.90$\times$                       \\
    1-FT Hx7Mesh      & 50176           & 657.2          & 0.06$\times$                       & 7.1                  & 0.91$\times$                       \\
    TPUv4             & 4096            & 185.7          & 0.22$\times$                       & 4.2                  & 5.52$\times$                       \\
    3D-Torus          & 4096            & 45.0           & 0.05$\times$                       & 4.2                  & 1.46$\times$                       \\
    2D 1-FT         & 4096            & 375.6          & 0.45$\times$                       & 50                   & 0.90$\times$                       \\
    \midrule
    RailX4Mesh        & 65536           & 751.1          & 0.06$\times$                       & 12.5                 & \textbf{0.45}$\boldsymbol{\times}$ \\
    RailX7Mesh        & \textbf{200704} & 1314.4         & \textbf{0.03}$\boldsymbol{\times}$ & 7.1                  & \textbf{0.45}$\boldsymbol{\times}$ \\
    \midrule
    4-Tier FT         & 196608          & 83718          & 2.10$\times$                       & 100                  & 2.10$\times$                       \\
    49:7:1 Tap. FT    & 200704          & 22052          & 0.54$\times$                       & 2.0                  & 26.5$\times$                       \\
    2-FT Hx7Mesh      & 200704          & 5822           & 0.14$\times$                       & 7.1                  & 2.01$\times$                       \\
    \bottomrule
  \end{tabular}
\end{table}

The results are shown in Table~\ref{tab:cost} (details are in \S~\ref{appendix:cost}~Table~\ref{appendix:tab:cost}), where the upper part is evaluated at maximum or original scale, and the lower part is evaluated at uniform scale ($\sim$200K chips). The cost per injection bandwidth approximates the cost per All-Reduce bandwidth. Compared with Torus, RailX uses high-bandwidth 2D-mesh to eliminate all the copper cables and significantly improve the bisection bandwidth. Compared with the Fat-Tree, RailX achieves the same All-Reduce bandwidth at less than $10\%$ cost. Compared with HammingMesh, which is the previous SOTA topology, RailX uses optical switching to reduce half of the cost while maintaining the throughput and improving the scalability.

\subsection{All-to-All Performance}
\label{sec:a2a-performance}
\begin{figure}[tb]
  \centering
  \includegraphics[width=0.99\linewidth]{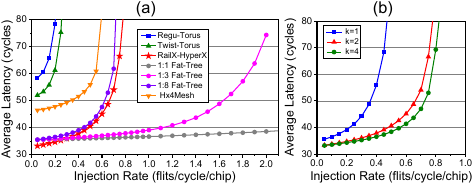}
  \caption{All-to-all performance. (a) Different topologies with ~1.3K chips; (b) \textit{RailX} with different intra-mesh bandwidth ($k$ times the inter-node bandwidth). \label{fig:all-to-all}}
\end{figure}

We simulate the all-to-all performance of \textit{RailX-2D-HyperX} ($m=4, n=2, N=1296$) and other topologies. All topologies have around 1.3K chips, and each chip has 8 ports (\textit{i.e.,} 8 flits/cycle/chip injection bandwidth). As shown in Figure~\ref{fig:all-to-all}(a), \textit{RailX} achieves 0.8 flits/cycle/chip throughput, which is close to the theoretical maximum throughput of 1 flits/cycle/chip. Compared with all other topologies, \textit{RailX} achieves the most cost-effective all-to-all throughput. Compared with HammingMesh, \textit{RailX} adopts higher intra-mesh bandwidth and eliminates the packet switches, thus achieving higher actual throughput and lower latency.

As analyzed in \S~\ref{sec:high-bandwidth-mesh}, the intra-mesh bandwidth is essential for \textit{RailX} as the 2D-mesh is utilized as a virtual switch. We simulate the performance with different internal bandwidths. As shown in Figure~\ref{fig:all-to-all}(b), if the internal bandwidth is the same as the external bandwidth, the performance is poor because the 2D-mesh becomes the bottleneck. With 2$\times$ internal bandwidth, the performance is significantly improved and close to the theoretical maximum throughput. Higher internal bandwidths do not bring significant improvement, as the external bandwidth becomes the bottleneck.

\subsection{All-Reduce Performance}
\label{sec:all-reduce-performane}
\begin{figure}[tb]
  \centering
  \includegraphics[width=0.99\linewidth]{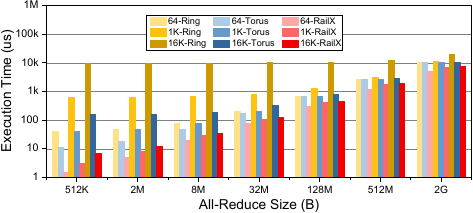}
  \caption{All-Reduce performance. \label{fig:all-reduce}}
\end{figure}

We also evaluate the All-Reduce performance of \textit{RailX}. 1D-ring, 2D-Torus~\cite{Hoefler_HammingMeshNetworkTopology_2022}, and hierarchical-2D-Torus algorithm presented in \S~\ref{sec:hierarchical-allreduce} are evaluated with different scales and sizes. We assume each chip has four ports (double bandwidth for the 1D-ring), the external bandwidth is set to $100$GB/s per port, and the internal bandwidth is set to $400$GB/s per port (\textit{i.e.,} $200$GB/s for 1D-ring All-Reduce on 2D-mesh). The latency is set to $300$ns per external hop~\cite{Sensi_SwingShortcuttingRings_2024} and $10$ns per internal hop.

As shown in Figure~\ref{fig:all-reduce}, the hierarchical-2D-Torus algorithm on RailX always achieves the best performance (shortest time to finish). For large-size All-Reduce, since all algorithms are near bandwidth-optimal, they achieve similar performance. For small-size All-Reduce, the 1D-ring algorithm has the worst performance, and the hierarchical-2D-Torus algorithm outperforms the 2D-Torus algorithm, especially at hyper-scale.

\subsection{Bandwidth Allocation Exploration}
\begin{figure}[tb]
  \centering
  \includegraphics[width=0.99\linewidth]{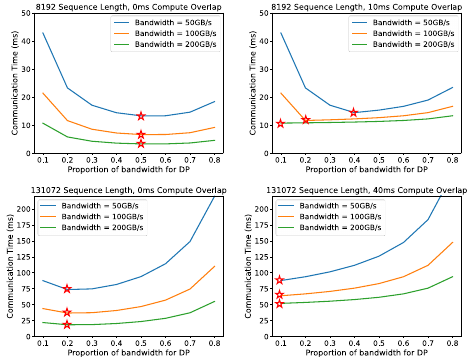}
  \caption{Different sequence length and computation-communication overlap introduce different bandwidth allocation strategies. \label{fig:static-allocation}}
\end{figure}

As discussed in \S~\ref{sec:bandwidth-allocation}, the bandwidth of one physical dimension is allocated to two parallelism dimensions according to the communication volume and the overlap with computation. The results of allocation (10 ports in total) between DP and CP with different sequence lengths are shown in Figure~\ref{fig:static-allocation}, where the actual communication times under different allocation proportions are measured. For small-scale sequence length, the CP communication is relatively small; therefore, the optimal policy is to allocate more bandwidth to DP. As the sequence length increases, the CP communication rises; thus, we tend to allocate more bandwidth to CP. After considering the computation/communication overlap of DP, we are supposed to allocate even more bandwidth to CP. In summary, optical switching with the \textit{dimension-splitting} method provides flexibility to adjust the bandwidth of each parallelism dimension for different configurations.

\label{sec:dimension-splitting-results}

\subsection{Reliability and Availability}

Similar to TPUv4~\cite{Zu_ResiliencyScaleManaging_2024}, \textit{RailX} also gains reliability and availability benefits from bypassing failure nodes with OCS configurability. However, due to the 2D organization, a failure node will cause the row and the column to be disconnected, which is similar to the \textit{HammingMesh}~\cite{Hoefler_HammingMeshNetworkTopology_2022}. By using proper allocation policy, we can fully utilize all nodes (as shown in \S~\ref{appendix:allocation-with-fault}) in \textit{ML-as-a-service (MLaaS)} scenario.

However, when using the entire system for one single job, there is no optimization room for utilizing disconnected nodes; therefore, the failure nodes will significantly affect availability. In the best case, all failure nodes are in the same row or column; we only lose one row or one column of nodes; in the worst case, the failure nodes (\textit{e.g.}, $2a$) are distributed in different rows and columns; as a result, the maximum allocation for single job is $(R/2-a)\times(R/2-a)$. For general cases, the maximum allocation is an NP-hard problem; however, since the failures are sparse, we can use a fast algorithm (shown in \S~\ref{appendix:allocation-with-fault}) to find the maximum allocation.

\begin{figure}[tb]
  \centering
  \includegraphics[width=0.97\linewidth]{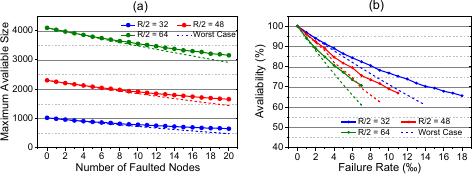}
  \caption{Availability of single allocation with OCS configurability. (a) Maximum size for a single job with faulted nodes; (b) Availability at failure rates. \label{fig:availability}}
\end{figure}

In evaluation, we randomly select nodes as failure nodes and calculate the maximum allocation for a single job. We use 100 random samples for each failure rate and calculate the average. As shown in Figure~\ref{fig:availability}(a), the maximum available size decreases linearly with the number of failure nodes, and the average size is significantly better than the worst case when the failure rate is high. As shown in Figure~\ref{fig:availability}(b), with the same failure rate, the large-scale \textit{RailX} has worse availability than the small-scale \textit{RailX}. Considering a typical failure rate of 0.1\%~\cite{Zu_ResiliencyScaleManaging_2024}, the availability is always more than 90\%, which is acceptable for large-scale training systems.

\section{Related Works}
\textit{HammingMesh} also leverages local 2D-mesh networks, providing high All-Reduce bandwidth at a low cost~{\cite{Hoefler_HammingMeshNetworkTopology_2022}}. From the topology perspective, \textit{RailX} is a circuit-switched \textit{HammingMesh} with higher local bandwidth inside the 2D-mesh. As a result, \textit{RailX} achieves higher scalability and higher AllReduce throughput than \textit{HammingMesh}. Limited by the space, more related works are shown in \S~\ref{appendix:related-work}.

\section{Conclusions}
\textit{RailX} is a novel network architecture optimized for large-scale LLM training systems. By utilizing advanced integration technologies (high-bandwidth 2D-mesh), 2D-organized circuit switches, and \textit{rail-ring-based all-to-all} interconnection, \textit{RailX} achieves higher scalability and cost-effectiveness than existing topologies. It provides extreme injection/All-Reduce bandwidth at hyper scales and very low cost (less than $10\%$ compared with the Fat-Tree) while maintaining sufficient bisection/All-to-All bandwidth. \textit{RailX} also demonstrates high flexibility in the MLaaS scenario. Circuit switching allows single or multiple jobs to be flexibly scheduled, and failure nodes can be easily worked around. With the \textit{dimension-splitting} method, the number of topology dimensions and the bandwidth/scale of each dimension can be flexibly adjusted according to the workload.


\bibliographystyle{ACM-Reference-Format}
\bibliography{reference}

\newpage
\appendix
\section{Appendices}
\label{sec:appendix}
\subsection{Hamiltonian Decomposition of Complete Graph}
\label{appendix:construction}
\begin{figure}[htb]
  \centering
  \includegraphics[width=0.99\linewidth]{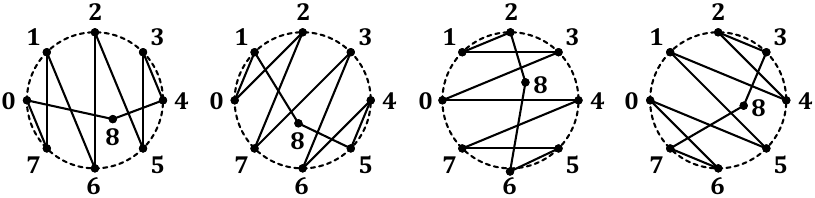}
  \caption{Hamiltonian decomposing of $K_{2m+1}^*$ complete graph.  \label{fig:hamiltonian-decomposing}}
\end{figure}
It has been proven that a directed complete graph $K_n (n\neq4,6)$ can always be decomposed into $n-1$ directed Hamiltonian cycles~\cite{Tillson_HamiltonianDecompositionK2m_1980,Alspach_WonderfulWaleckiConstruction_2008}. If $n=2m+1$, there is a straightforward construction to decompose the graph into $m$ undirected (bidirectional) Hamiltonian cycles. As shown in Figure~\ref{fig:hamiltonian-decomposing}, first, $m$ Hamiltonian paths are constructed among $2m$ vertices by
\begin{equation*}
  (i, i-1, i+1, i-2, i+2, \cdots, i+m-1, i-m)\mod{2m},
\end{equation*}
where $i=0, ..., m-1$. Then, the two endpoints of each path are connected to the vertice $2m+1$, forming a cycle. We can see that each cycle is a Hamiltonian cycle, and there is no overlap between cycles under this construction.

If $n=2m$, the graph can be decomposed into $2m-1$ directed Hamiltonian cycles.
The construction is present in ~\cite{Alspach_WonderfulWaleckiConstruction_2008}, which is relatively complicated and will not be illustrated here.

\begin{figure*}[!htb]
  \centering
  \includegraphics[width=0.97\linewidth]{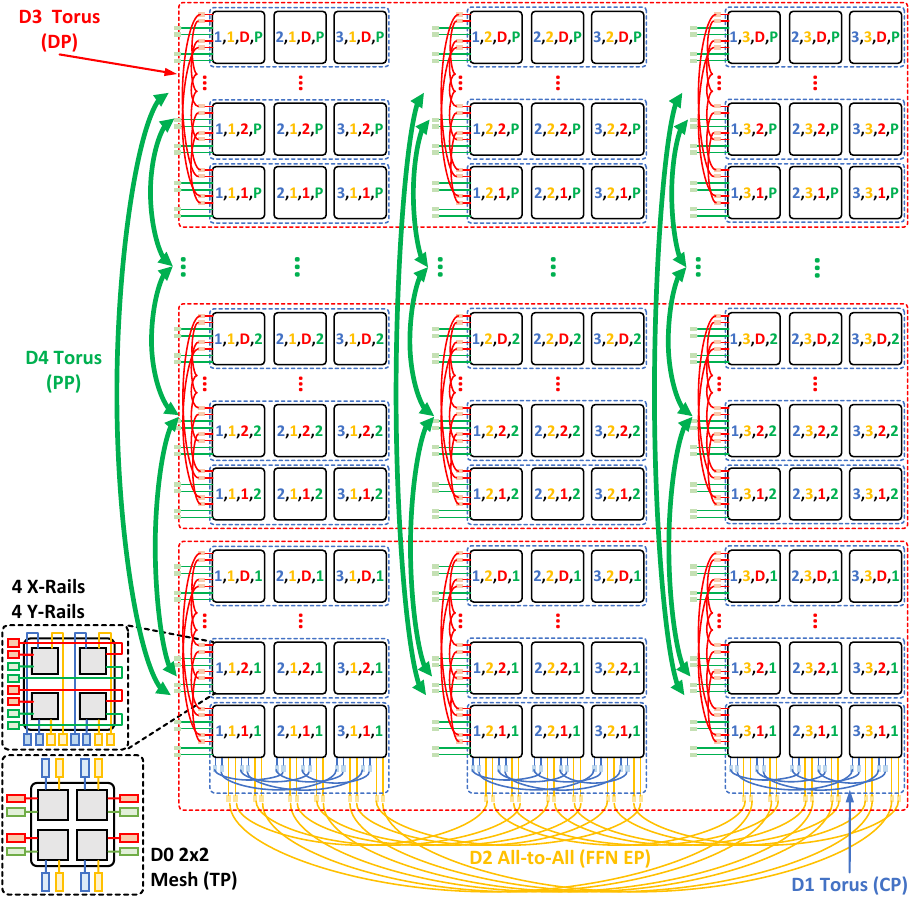}
  \caption{High-dimensional heterogeneous topology ($4\times 3 \times 3 \times D \times P$) and the mapping of the training workload ($TP \times CP \times EP \times DP \times PP$).\label{fig:example}}
\end{figure*}

\subsection{All-to-All-based All-Reduce}
\label{appendix:a2a-allreduce}
For HyperX configuration, we can perform direct reduce-scatter and all-gather algorithms on RailX based on the all-to-all interconnection. That is to say, each node sends/receives data to every other node simultaneously. The communication time of the 2D-HyperX All-Reduce algorithm is estimated as
\begin{equation}
  T_{AR} (p, V, B) = \alpha + \frac{p-1}{p}\frac{V}{2B},
\end{equation}
\begin{equation}  
  \begin{aligned} 
    T_\text{2D-HyperX} =\;& 2\times\frac{m^2-1}{m^2}\frac{V}{2\boldsymbol{k}nB} (\textit{All-Reduce on local 2D-mesh}) \\ 
     &+ 2\left[T_{AR}(p, \frac{V}{2\boldsymbol{m^2}}, \frac{nB}{\boldsymbol{m}})+T_{AR}(mp, \frac{V}{2\boldsymbol{m^2}p}, \frac{nB}{\boldsymbol{m}})\right] \\[-2pt]
                               =\;& \frac{m^2-1}{m^2}\frac{V}{\boldsymbol{k}nB} + 4\alpha + \frac{p^2-1}{p^2}\frac{V/\boldsymbol{m^2}}{2nB/\boldsymbol{m}}              \\[-2pt]
                               \approx\;& 4\alpha + \left(\frac{2}{k}+\frac{1}{m}\right)\frac{V}{2nB},
  \end{aligned}
\end{equation}
which does not increase with the scale $p$.
  
\subsection{High-Dimensional Heterogeneous Topology and Workload Mapping}
\label{appendix:example}
A detailed example of mapping 5D parallelism on RailX is shown in Figure~\ref{appendix:example}. Tensor parallelism is mapped on the high-bandwidth 2D-mesh. Rails for expert parallelism are configured into all-to-all topology, and other rails are configured into 3D-Torus.

\begin{table*}[tb]
  \fontsize{8pt}{10pt}\selectfont
  \setlength{\tabcolsep}{3pt}
  \renewcommand\arraystretch{1.2}
  \caption{Communication scope, volume, and frequency of high-dimensional parallelism on RailX \label{tab:communication}}
  \begin{tabular}{|r|r|l|c|c|c|}
    \hline
    Parallelism                           & Traffic Pattern                                 & Process Group                                                                         & Volume ($V$)                  & Frequency (F) & Communication Time                      \\
    \hline
    \multirow{2}{*}{Tensor\,/\,SEQ ($T$)} & \multirow{2}{*}{\makecell[{{r}}]{Reduce-Scatter                                                                                                                                                                                   \\ \& All-Gather}}       & $G_{T, i}=\{i, i+1, ..., i+T_a-1\}$ (Attention Blocks)                                                   & $BSH$                    & $4N_BL/P$            & $V/2kn \times F$                        \\
    \cline{3-6}
                                          &                                                 & $G_{T, i}=\{i, i+1, ..., i+T_e-1\}$ (FFN/Expert Blocks)                               & $BSHK$                        & $4N_BL/P$     & $V/2kn \times F$                        \\
    \hline
    \hline
    Context ($C$)                         & Point-to-Point                                  & $G_{C, i}=\{i, i+T, ..., i+(C-1)T\}$                                                  & $BSH\times (2h_{KV}/h_{A})/T$ & $2N_BL/P$     & $(C\alpha + V/\frac{2n_c}{m}) \times F$ \\
    \hline
    Expert ($E$)                          & All-to-All                                      & $G_{E, i} = \{i, i+TC, ..., i+(E-1)TC\}$                                              & $BSHK/TC$                     & $4N_BL/P$     & $(\alpha + V/\frac{2n_e}{m}) \times F$  \\
    \hline
    \hline
    \multirow{3}{*}{Data ($D=ED_e$)}      & All-Reduce                                      & $G_{D_{VOC}, i}=\{i, i+TC, ..., i+(D-1)TC\}$                                          & $2HV / TC$                    & $1$           & $ T_{2D}(n_e, n_d) \times F$            \\
    \cline{2-6}
                                          & All-Reduce                                      & $G_{D_\text{QKV}, i}=\{i, i+T, ..., i+(CD-1)T\}$                                      & $(2+2h_{KV}/h_A)H^2 / T$      & $L/P$         & $ T_{3D}(n_c, n_e, n_d) \times F$       \\
    \cline{2-6}

                                          & All-Reduce                                      & $G_{D_\text{FFN},i}=\{G_{C, i}, G_{C, {i+TCE}},..., G_{C, {i+(D_e-1)TCE}}\}$          & $3HI / T$                     & $L/P$         & $T_{2D}(n_c, n_d) \times F$             \\
    \hline
    \hline
    Pipeline ($P$)                        & Point-to-Point                                  & $G_{P, i}=\{i \Leftrightarrow i+TCD \Leftrightarrow ... \Leftrightarrow i+(P-1)TCD\}$ & $BSH/TC$                      & $2N_B$        & $V/\frac{n_p}{m} \times F$              \\
    \hline
  \end{tabular}
\end{table*}

The communication pattern, volume, frequency, and communication time of each parallelism are shown in Table~\ref{tab:communication}, where $L$ is the layer number, $B$ is the micro batch size, $N_B$ is the number of micro batches per DP (global batch size $= B \times N_B \times D$), $S$ is the sequence length, $H$ is the hidden dimension, $V$ is the vocabulary size, $h_A$ and $h_{KV}$ is the attention and key/value head number (grouped-query attention~\cite{Ainslie_GQATrainingGeneralized_2023}), $I$ is the intermediate dimension of FFN layers(SwiGLU activation function~\cite{Shazeer_GLUVariantsImprove_2020}), and $K$ is the Top-$K$ gating used for the MOE structure~\cite{Shazeer_OutrageouslyLargeNeural_2017}. The communication time is estimated as the sum of the latency and bandwidth part, where $\alpha$ is the latency, $T_{2D}$ and $T_{3D}$ are the communication time of 2D and 3D All-Reduce.

\subsection{HW/SW Configuration and Traces}
\label{appendix:trace}
To validate the analytical model, we run training workloads on real hardware and collect the trace by using NVIDIA Nsight Systems. Up to 64 NVIDIA GH200 chips~\cite{NVIDIA_NVIDIAGraceHopper_2024} with 96 GB HBM3 and 120 GB LPDDRX5 are used. Each scale-up node consists of four GH200 chips, all-to-all interconnected with 1.2Tb/s NVLink4.0 (six 200Gb/s links). All nodes are interconnected in a Slingshot-11~\cite{DeSensi_InDepthAnalysisSlingshot_2020, DeSensi_ExploringGPUtoGPUCommunication_2024} network with 200Gb/s NIC per each GH200 chip. Megatron-LM 0.9 (main-branch commit-7efaa73), Pytorch 3.12, NCCL 2.23.4, and CUDA 12.6 are used.

\begin{lstlisting}[language=python,caption={Parallelism and MOE configurations},
  label={lst:parallelism_moe},basicstyle=\ttfamily\small]
# Parallelism Configuration
data_parallel_size: 4
tensor_model_parallel_size: 4
pipeline_model_parallel_size: 2
context_parallel_size: 2
expert_model_parallel_size: 2

# MoE Configuration
num_experts: 8
moe_router_load_balancing_type: aux_loss
moe_aux_loss_coeff: 0.01
moe_router_topk: 2
moe_token_dispatcher_type: alltoall
moe_grouped_gemm: True
\end{lstlisting}

\begin{figure*}[htb]
  \centering
  \includegraphics[width=0.99\linewidth]{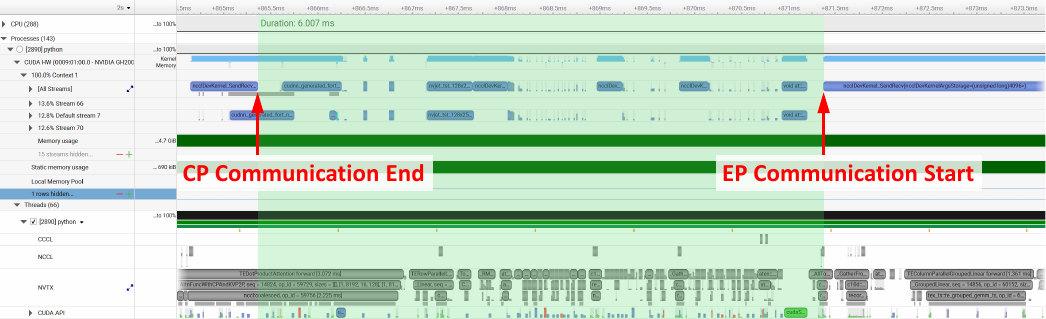}
  \caption{\label{fig:trace}Trace example. CP and EP are separated by 6 ms.}
\end{figure*}

An example trace of Llama3-70B training workload is shown in Figure~\ref{fig:trace}, and the related configurations are listed in Listing~\ref{lst:parallelism_moe}. From the trace, we can count the actual communication and computation time of each parallelism.

It can be measured that all-to-all communication starts 6ms after the context parallelism communication. Therefore, it is possible to configure the circuit switches in the gap between the context parallelism communication and the expert parallelism communication. Specifically, all ports of one physical dimension are configured to connect within the CP group during the CP communication and then switch to connect within the EP group before EP communication starts. As a result, both CP and EP communications can fully utilize the bandwidth of one entire physical dimension.

\subsection{Allocating Jobs on RailX with Faulted Nodes}
\label{appendix:allocation-with-fault}

\begin{figure}[h]
  \centering
  \includegraphics[width=0.99\linewidth]{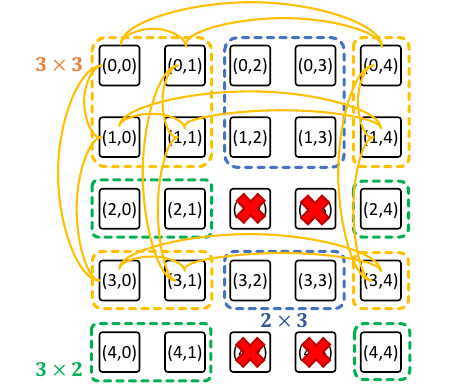}
  \caption{Allocation of jobs with failure nodes.  \label{fig:allocation-with-fault}}
\end{figure}

In the MLaaS scenario, we can use multiple small workloads to fully utilize all functional nodes in the faulted \textit{RailX}. As shown in Figure~\ref{fig:allocation-with-fault}, four faulted nodes are marked with red crosses. If using \textit{RailX} for a single job, two entire rows or columns of nodes must be disconnected to work around the faulted nodes. However, when multiple small jobs are allocated, three sub-networks fully utilize all the functional nodes. 

\begin{algorithm}[tbh]
  \caption{\scshape Maximum Availability in Faulted Grid \label{alg:max-allocation}}
  \begin{algorithmic}[1]
    \REQUIRE System size: $n\times n$; \\
    \quad Faulted nodes: $F = \{(r_i, c_i)\}$.
    \ENSURE Maximum available single-allocation size.
    \STATE $I \gets$ Isolated faulted nodes
    \STATE $C = F-I \gets$ Non-isolated faulted nodes
    \IF[all nodes are isolated]{$C$ is $\varnothing$}
    \STATE // \textit{evenly disable rows and columns}
    \RETURN $(n - ceil(|F|/2)) \times (n - floor(|F|/2))$
    \ENDIF

    \STATE $S_\text{max} \gets 0$
    \STATE // Iterate over all cases for $C$ (0/1: disable row/column)
    \FOR{each $\in \{0, 1\}^C$}
    \STATE $r_i \gets$ disabled rows
    \STATE $c_i \gets$ disabled columns
    \STATE Adjust $|I| = r' + c'$ ($r', c' \in N$)
    \STATE $S \gets \mathop{\text{max}}\limits_{r', c'}[(n - r_i- r') \times (n - c_i - c')]$
    \STATE $S_\text{max} \gets \max(S_\text{max}, S)$
    \ENDFOR
    \RETURN $S_\text{max}$
  \end{algorithmic}
\end{algorithm}

The general allocation has been proven to be a NP-hard problem~\cite{Hoefler_HammingMeshNetworkTopology_2022}. However, since the faulted nodes are sparse, which means they rarely appear in the same row or column, we can use a fast algorithm to find the maximum available single-allocation size. As shown in Algorithm~\ref{alg:max-allocation}, we first find all non-isolated faulted nodes (appearing in the same row or column with other faulted nodes) and isolated faulted nodes. Then, we iterate over all cases for non-isolated faulted nodes to disable rows or columns by complexity $O(2^{|C|})$; for each case, we maximize the available size by making width and height as close as possible. Finally, we return the maximum available size. Since the failure rate is low, faulted nodes are sparse, so $|C|$ is supposed to be small, and the algorithm is efficient.

\subsection{Chiplet Network Simulator}
\label{appendix:cnsim}
A sophisticated 2D-mesh is adopted within each node, whose performance is tightly linked to microarchitecture and routing algorithms. Fine-grained network simulation can be very slow~\cite{Jiang_DetailedFlexibleCycleaccurate_2013, Krishna_Garnet20DetailedOnChip_2017} (days to simulate more than 1K chips). \textit{Chiplet Network Simulator (CNSim)} is a recent cycle-accurate simulator that supports multi-threading simulation~\cite{Feng_EvaluatingChipletbasedLargeScale_2024}.

What we model by using CNSim includes router pipelines, virtual cut-through packet switching, fine-grained VC-based deterministic/adaptive routing algorithms, and non-uniform link bandwidth/latency. To speed up the simulation, we enable multi-threading, omit network protocols (\textit{e.g.}, TCP/IP), and use simplified link delay. The default parameters used in simulations are shown in TABLE~\ref{table:parameter}.

\begin{table}[h]
  \centering
  \setlength{\tabcolsep}{5pt}
  \renewcommand\arraystretch{1.1}
  \caption{Default Parameters}
  \label{table:parameter}
  \begin{tabular}{|l|l|}
    \hline
    \textbf{Parameter}     & \textbf{Value}                 \\
    \hline
    Packet Length          & $4$ flits                      \\
    \hline
    Input Buffer Size      & $16$ flits                     \\
    \hline
    Base Link Bandwidth    & $1$ flit/cycle                 \\
    \hline
    Short-Reach Link Delay & $1$ cycle                      \\
    \hline
    Long-Reach Link Delay  & $10$ cycles                     \\
    \hline
    Simulation Time        & $10000$ cycles  \vspace{-1pt}  \\
                           & after $5000$ cycles warming up \\
    \hline
  \end{tabular}
\end{table}

\subsection{Cost}
\label{appendix:cost}
\begin{table*}
  \centering
  \caption{Cost comparison with 64-port packet switches and 128-port circuit switches. Each chip has 36 ports. \label{appendix:tab:cost}}
  \begin{tabular}{lcccccccc}
    \toprule
    \textbf{Topology} & \textbf{Scale} & \textbf{Switch} & \textbf{PCC}    & \textbf{AOT}    & \textbf{Cost}  & \textbf{Cost}                      & \textbf{Glob. BW}    & \textbf{Cost}                      \\
                      & \textbf{[\#]}  & \textbf{[\#]}   & \textbf{[\#K]} & \textbf{[\#K]} & \textbf{[M\$]} & \textbf{[\,/\,Inject]}             & \textbf{[\% Inject]} & \textbf{[\,/\,GBW]}                \\
    \midrule
    2-Tier Nonbl. FT         & 2048           & 3456            & 0               & 294.9            & 415.9          & 1$\times$                        & 100                  & 1$\times$                        \\
    1:3 Tap. 2-Tier FT       & 3072           & 2880             & 0               & 294.9            & 395.7          & 0.65$\times$                       & 33.3                 & 1.90$\times$                       \\
    Hx4Mesh (1-Tier FT)         & 16384          & 2304             & 0               & 294.9            & 375.6          & 0.11$\times$                       & 12.5                 & 0.90$\times$                       \\
    Hx7Mesh (1-Tier FT)          & 50176          & 4032            & 0               & 516.1            & 657.2          & 0.06$\times$                       & 7.1                 &0.91$\times$                       \\
    3D-Torus w/ OCS            & 4096           & 288             & 30.7            & 36.9            & 185.7           & 0.22$\times$                       & 4.2                  & 5.52$\times$                       \\
    3D Torus w/o OCS       & 4096           & 0               & 30.7            & 36.9            & 45.0           & 0.05$\times$                       & 4.2                  & 1.46$\times$                       \\
    Rail-Only (2D FT)         & 4096 & 2304               &   0          & 294.9               & 375.6           & 0.45$\times$                      & 50                & 0.90$\times$                       \\
    \midrule
    RailX4Mesh        & 65536 & 4608            & 0               & 589.8           & 751.1          & 0.06$\times$ & 12.5                 & \textbf{0.45}$\boldsymbol{\times}$ \\
    RailX7Mesh        & \textbf{200704} & 8064           & 0               & 1032.2           & 1314.4          & \textbf{0.03}$\boldsymbol{\times}$ & 7.1                 & \textbf{0.45}$\boldsymbol{\times}$ \\
    \midrule
    4-Tier Nonbl. FT         & 196608           & 774144           & 0               & 56623           & 83718          & 2.10$\times$                        & 100                  & 2.10$\times$                       \\
    1:7:49 Tap. 3-Tier FT     & 200704          & 149760            & 0               & 16810           & 22052          & 0.54$\times$                       & 2.0                 & 26.5$\times$                       \\
    Hx7Mesh (2-Tier FT)    & 200704          & 48384           & 0               & 4128           & 5822          & 0.14$\times$                       & 7.1                 & 2.01$\times$                       \\
    \bottomrule
  \end{tabular}
\end{table*}
Detailed cost results are shown in Table~\ref{appendix:tab:cost}, which includes the number of switches, passive copper cables (PCCs), and active optical transceivers (AOTs). Since all topologies can achieve near-bandwidth-optimal All-Reduce with proper algorithms, the cost per injection bandwidth approximates the cost per All-Reduce bandwidth. With 64-port packet switches, 2-tier non-blocking Fat-Tree can only connect 2048 chips, the cost of which is regarded as baseline. By over-subscribing the switches, 3:1 Taper Fat-Tree can connect 3072 chips at a lower cost, but the global bisection bandwidth is only one-third of the non-blocking Fat-Tree, so the cost per global bandwidth is higher. With 128-port circuit switches, the cost of which is assumed to be the same as the 64-port packet switch, RailX4Mesh can connect 65536 chips with 12.5\% of the bisection bandwidth, and RailX7Mesh can connect 200704 chips with 7.1\% of the bisection bandwidth. As a result, \textit{RailX} achieves much higher scalability with a much lower cost per injection bandwidth while maintaining a similar cost per global bandwidth.

To achieve the same scale as the 128-port-OCS-based RailX7Mesh ($\sim 200$K chips), four tiers of switches are required, costing more than two times the 2-tier Fat-Tree. By using the 49:7:1 Taper Fat-Tree, we can reduce the number of tiers to three, but the cost per global bisection bandwidth will be significantly higher.

\subsection{Related Work}
\label{appendix:related-work}

\textit{BML}~{\cite{Wang_ScalableHighPerformanceFaultTolerant_2020}} is a network architecture based on the BCube topology~{\cite{Guo_BCubeHighPerformance_2009}}. It achieves higher All-Reduce bandwidth with fewer switches than the traditional Fat-Tree-based parameter server. However, only data parallelism is considered in BML, which is insufficient for the modern hyper-scale LLM training workload. 

\textit{TopoOpt}~\cite{Wang_TopoOptCooptimizingNetwork_2023} and \textit{SiP-ML}~\cite{Khani_SiPMLHighbandwidthOptical_2021} are two recent reconfigurable network architectures for ML training. They both use commercially available optical circuit switches and online/offline methods to optimize the network topology for specific workloads. However, their scalability is limited by the radix of optical switches, and they are not optimized for all-to-all communication, which is essential for MOE models.

\textit{Rail-Only} removes all cross-rail switches in traditional rail-optimized Fat-Tree and uses a scale-up network for cross-rail communication. From the topology perspective, \textit{Rail-only} is a 2D Fat-Tree (scale-up and scale-out). In \S~\ref{sec:cost-analysis}, we let the two Fat-Tree networks have the same bandwidth because this leads to the best global All-to-All throughput.

\end{document}